\documentclass[article,nofootinbib,superscriptaddress]{revtex4}

\usepackage{hyperref}
\usepackage{epigraph}
\usepackage{hyperref}
\usepackage{latexsym, graphics,multirow,comment,appendix} 

\usepackage{amsfonts}
\usepackage{bm}
\usepackage{color}
\usepackage{graphicx}
\usepackage{amsmath}
\usepackage{amsfonts}
\usepackage{amssymb}
\usepackage{color}
\usepackage{epsf}
\usepackage[a4paper]{geometry} 
\usepackage{slashed}
\usepackage{mathtools}

\newcommand{\be}{\begin{eqnarray}}
\newcommand{\ee}{\end{eqnarray}}
\newcommand{\ba}{\begin{eqnarray}}
\newcommand{\ea}{\end{eqnarray}}

\newcommand{\no}{\nonumber}

\newcommand{\G}{\mathcal{G}}

\begin{document}

\vspace*{-30mm}

\title{\boldmath High energy neutrino telescopes as a probe of the neutrino mass mechanism}

\author{Kfir Blum}\email{kblum@ias.edu}
\author{Anson Hook}\email{hook@ias.edu}
\author{Kohta Murase}\email{murase@ias.edu}
\affiliation{School of Natural Sciences, Institute for Advanced Study, Princeton, New Jersey 08540, USA}

\vspace*{1cm}

\begin{abstract} 
We show that measurements of the spectral shape and flavor ratios of high energy astrophysical neutrinos at neutrino telescopes can be sensitive to the details of the neutrino mass mechanism.  We propose a simple model for  Majorana neutrino mass generation that realizes the relevant parameter space, in which small explicit lepton number violation is mediated to the Standard Model through the interactions of a light scalar.  IceCube, with about ten years of exposure time, could reveal the presence of anomalous neutrino self-interactions. 
Precision electroweak and lepton flavor laboratory experiments and a determination of the total neutrino mass from cosmology would provide consistency checks on the interpretation of a signal. 
\end{abstract}

\maketitle

%\tableofcontents
%============================================================================================================================
\section{Introduction}
\label{sec:intro}

The IceCube collaboration has recently reported the first measurement of an extraterrestrial high energy neutrino flux~\cite{Aartsen:2013bka,Aartsen:2013jdh,Aartsen:2014gkd}.  
The observed flux is consistent with isotropic arrival distribution and an equal mix of $\nu_e,\nu_\mu,\nu_\tau$ with equal number of neutrinos and antineutrinos.  The spectral slope is consistent with constant power per log energy bin ($\epsilon_\nu^2 J_\nu\sim$~const), with a possible suppression above a few PeV.    
The normalization is consistent with the Waxman-Bahcall bound within uncertainty in source-redshift evolution~\cite{Waxman:1998yy,Bahcall:1999yr}.  Though statistics are still limited, all of these factors together hint to a cosmological origin for the observed neutrino flux, most likely tied to the sources of high energy cosmic rays (for reviews see, e.g.~\cite{Waxman:2013zda,Halzen:2013dva,Meszaros:2014tta}).  The cosmological origin is also supported by gamma ray data
~\cite{Ahlers:2013xia,Murase:2013rfa}, although a subdominant Galactic contribution is not yet excluded.   

While the basic promise of high energy neutrino detectors is to help unravel the accelerators of high energy cosmic rays, it is interesting to contemplate in addition whether the new measurement could have implications for fundamental particle physics. One direction could be to hypothesize some new physics source for the neutrino flux, such as dark matter decay~\cite{Murase:2012xs,Feldstein:2013kka,Esmaili:2013gha,Bai:2013nga,Ema:2013nda}. This would make the broad consistency of the observed neutrino flux with the energy budget of high energy cosmic rays (see e.g.~\cite{Murase:2013rfa,Katz:2013ooa}) an accident. 
Another possibility, that we find more appealing, is to use astrophysical neutrinos as a probe of new physics, and contemplate whether high energy neutrinos could experience anomalous interactions during their cosmologically long journey from the astrophysical source to IceCube.  From the particle physics perspective, it is important to remember that the neutrino sector holds a mystery.  We do not know what mechanism generates neutrino masses, but whatever this mechanism is, it is not part of the standard model (SM). To some degree, neutrino interactions beyond the SM are guaranteed.

In this paper we propose a model for Majorana neutrino mass generation that results in distortions of the flux and flavor composition of high energy neutrinos, measurable in neutrino telescopes. We show that an interesting parameter space exists in which IceCube could provide the first hints to the existence and structure of low-scale neutrino mass generation. 

Neutrino self-interactions were analyzed extensively in the literature. Refs.~\cite{Chacko:2003dt,Hall:2004yg,Davoudiasl:2005ks,Friedland:2007vv} discussed the relation to low-scale neutrino mass generation, focusing on spontaneously broken global and gauged lepton number.  
As we explain in Sec.~\ref{sec:model}, interesting phenomenology in neutrino telescopes leads us to consider explicit rather than spontaneous breaking of lepton number, implying a twist in model-building compared to Refs.~\cite{Chacko:2003dt,Hall:2004yg,Davoudiasl:2005ks,Friedland:2007vv}.  
We demonstrate how small explicit lepton number violation could be combined with a low-scale mechanism for neutrino masses.  While this scenario is, in some respects, less predictive than the case of spontaneous symmetry breaking, it is simple, technically natural and opens the way to new phenomenology in the neutrino sector.

An analysis closely related to ours was presented in~\cite{Goldberg:2005yw,Baker:2006gm}, which studied the effect of light scalar exchange on the energy spectrum of $\sim$10~MeV neutrinos from core-collapse supernovae.  Effects due to vector boson exchange on the neutrino flux at high energy neutrino telescopes were considered in~\cite{Keranen:1997gz,Hooper:2007jr}.  More recently, Refs.~\cite{Ng:2014pca,Ioka:2014kca} presented IceCube constraints on neutrino interactions through a light mediator.  In contrast to these works, we explore a concrete model with a well defined relation to the neutrino mass mechanism.  This allows us to (i) analyze neutrino flavor effects, highlighting the interplay between the rich phenomenology of a three-flavor detection at IceCube to the flavor structure governing neutrino oscillations; and (ii) contrast our model with concrete experimental constraints. 

Many constraints on neutrino self-interactions were derived in the literature based on laboratory, astrophysical and cosmological data. We recalculate the most relevant constraints and refer to the corresponding literature in the body of the paper. 

The paper is organized as follows.  
In Sec.~\ref{sec:model} we write an effective Lagrangian for neutrino masses including a light scalar $\Phi$. We identify the parameter space that is relevant for high energy neutrino astronomy, where high energy astrophysical neutrinos scatter on the ambient cosmic neutrino background (C$\nu$B) through resonant $\Phi$ particle exchange. We then propose a simple model that realizes this parameter space using heavy Dirac sterile neutrinos and explicit breaking of lepton number mediated to the SM through the interactions of $\Phi$.  
In Sec.~\ref{sec:nutele} we calculate the effects of the neutrino interactions on the spectrum and flavor composition observable at neutrino telescopes. We highlight the relation between the  spectral and flavor distortions to the details of the neutrino mass mechanism.  
We assess the prospects for detection by calculating neutrino event rates in the IceCube detector, considering both showers and tracks.   
In Sec.~\ref{sec:disc} we summarize our results.  
In App.~\ref{app:xs} we collect formulae for neutrino self-interactions.  
In App.~\ref{app:expcst} we summarize observational constraints including meson decay, neutrinoless double-beta decay, electroweak precision tests, lepton flavor violation, as well as astrophysical and cosmological constraints.

%%%%%%%%%%%%%%%%%%%%%%%%%%%%%%%%
\section{Low-scale neutrino masses with neutrino self-interactions}
\label{sec:model}

Consider the low energy effective Lagrangian describing neutrino mass generation
\be\label{eq:L1}\mathcal{L}&=&-\frac{g}{\Lambda^2}\Phi(HL)^2+cc,\ee
where $\Lambda$ is a large mass scale, $g$ is a dimensionless coupling (matrix in lepton flavor), and $\Phi$ is a SM-singlet complex scalar. 
We work in Unitary gauge, where electroweak symmetry breaking is described by $H=\frac{1}{\sqrt{2}}(0\;\;v+h)^T$ with $v=246$~GeV. $L=(\nu\;\;l^-)^T$ is the SM lepton doublet left-handed Weyl spinor, and we denote the antisymmetric SU(2) contraction by $(HL)=H^Ti\sigma^2L$.
Lepton number violation is mediated to the SM through a vacuum expectation value for $\Phi$,
\be\label{eq:mu}\Phi=\phi+\mu\ee
with $\langle\Phi\rangle=\mu$. 
In the neutrino mass basis we have
\be\label{eq:flav}\mathcal{L}&=&-\frac{1}{2}\sum_i\left(m_{\nu_i}+\G_i\phi\right)\nu_i\nu_i+cc+...,\ee
with
\be m_{\nu_i}=\frac{g_i\mu v^2}{\Lambda^2},\;\;\;\;\;g={\rm diag}(g_1,g_2,g_3),\;\;\;\;\;\G_{i}=\frac{m_{\nu_i}}{\mu}=\frac{g_iv^2}{\Lambda^2}\ee
and where the $...$ in Eq.~(\ref{eq:flav}) stand for Higgs interactions that we do not discuss here. For later convenience we define
\be\label{eq:Gdef}\G\equiv\sum_i\G_i=\frac{\sum_i m_{\nu_i}}{\mu}.\ee

Focusing our attention to the phenomenology at neutrino telescopes, we show later on in Sec.~\ref{ssec:bol} that a sizable modification to the neutrino flux observed at earth occurs if
\be\label{eq:Gbig}\G\gtrsim10^{-3}\left(\frac{m_\phi}{10~\rm MeV}\right),\;\;\;\;\;{\rm or\;equivalently}\;\;\;\;\;\Lambda\lesssim 8~{\rm TeV}\times\left(\frac{m_\phi}{10~\rm MeV}\right)^{-\frac{1}{2}}g^{\frac{1}{2}}.\ee
The main observable effect is the scattering of high energy neutrinos on C$\nu $B through resonant $\phi$ exchange, with resonance energy 
\be\label{eq:Eres}\epsilon_{\rm res}=\frac{m_\phi^2}{2m_{\nu}}=1~{\rm PeV}\left(\frac{m_\phi}{10~{\rm MeV}}\right)^2\left(\frac{m_{\nu}}{0.05~{\rm eV}}\right)^{-1}.\ee
For the scattering to be identifiable in a neutrino telescope of the scale size of IceCube, the resonance energy should fall in the range between a few TeV to a few PeV, where the atmospheric background becomes manageable but the statistics is still large enough for a reasonable exposure time. 
Note that the scattering effect persists somewhat below $\epsilon_{\rm res}$, since the resonance energy of neutrinos from high-redshift sources is lower by $1+z$ as seen at the Earth.  Non-resonant interactions can in principle be important for large values of $\mathcal G$~\cite{Ng:2014pca,Ioka:2014kca}, but we show that such large values are excluded in our model by various experiments.  

\begin{figure}[!h]\begin{center}
\includegraphics[width=0.7\textwidth]{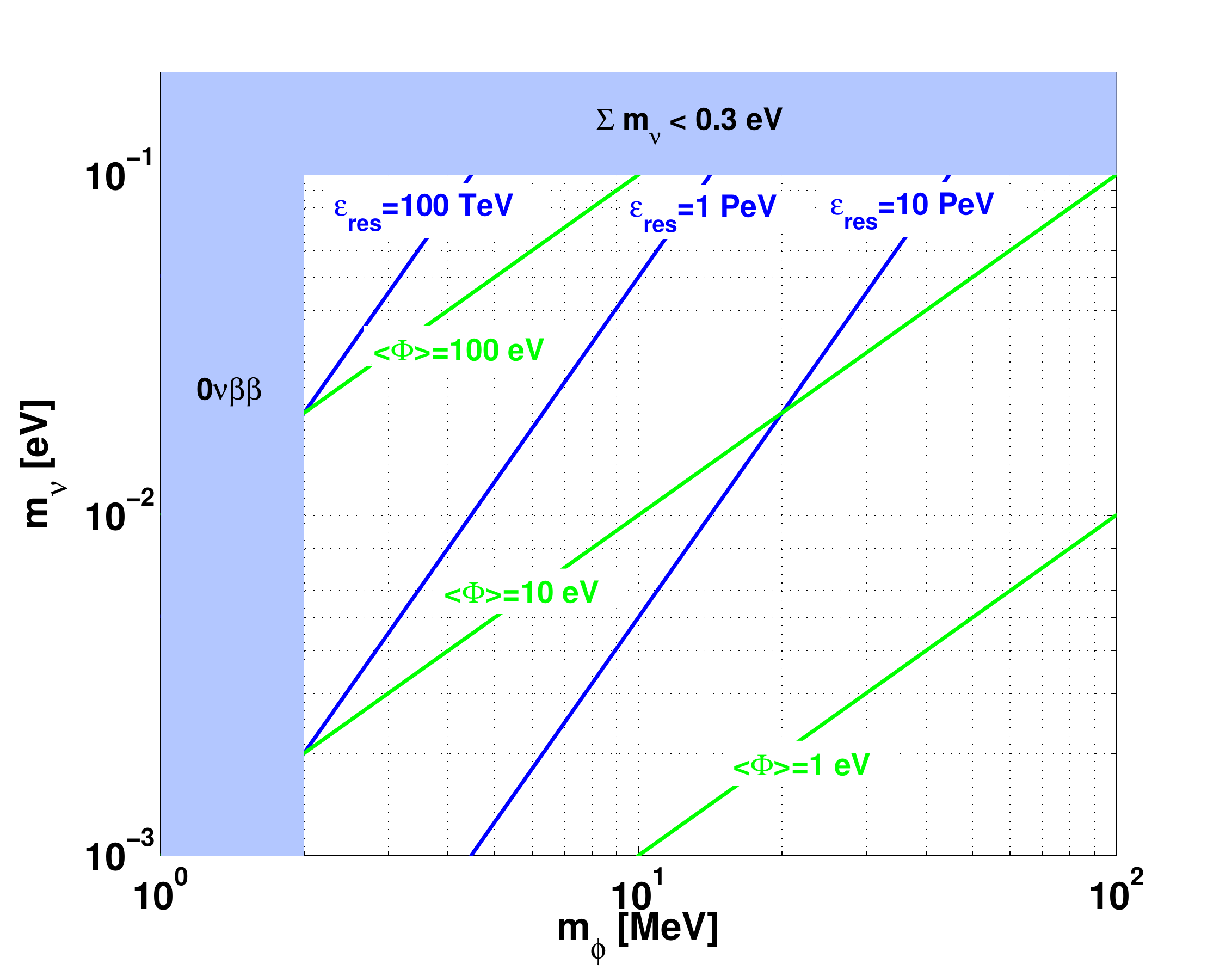}\end{center}
\caption{Parameter space for neutrino-neutrino scattering with visible effects at high energy neutrino telescopes of size $\sim$km$^3$. The x-axis gives the scalar mass and the y-axis the neutrino mass. Blue contours show the resonance energy, Eq.~(\ref{eq:Eres}), where an absorption feature in the neutrino spectrum shows up. For detectability at a neutrino telescope in a reasonable time scale (order ten years), one would need $\epsilon_{\rm res}\sim10-1000$~TeV. Green contours show the maximal value for the scalar VEV $\langle\Phi\rangle\equiv\mu$, deduced from Eq.~(\ref{eq:mureq}) to give an appreciable optical depth for neutrino-neutrino scattering over cosmological distance. The blue-shaded region on the left is excluded by limits on majoron-emitting neutrinoless double-beta decay (see App.~\ref{app:expcst}), and on the top is disfavored by cosmological constraints on the sum of neutrino masses~\cite{Hinshaw:2012aka,Ade:2013zuv,Mantz:2014paa}. 
}
\label{fig:ps}
\end{figure}

There are then two basic requirements on the new physics leading to Eq.~(\ref{eq:L1}):
%%%%
\begin{enumerate}
\item Requiring $\epsilon_{\rm res}\sim$TeV-PeV and using Eq.~(\ref{eq:Gbig}), we find that the new physics scale needs to be quite close to the electroweak scale, $\Lambda=\mathcal{O}\left(10~{\rm TeV}\right)$.  
\item Eq.~(\ref{eq:Gbig}) implies
\be\label{eq:mureq}  \mu\lesssim \left(\frac{m_\phi}{10~\rm MeV}\right)^{-1}\left(\frac{\sum_i m_{\nu_i}}{0.1~\rm eV}\right)~100~{\rm eV}.\ee
We thus need to explain a large gap between the scalar mass and its Vacuum Expectation Value (VEV): $m_\phi\gg\langle\Phi\rangle=\mu$. Explaining such a gap would be difficult if lepton number was broken spontaneously by $\Phi$. The lesson we take from this constraint is that lepton number violation should be explicit in the $\Phi$ sector. 
\end{enumerate}
%%%
Considering effects in neutrino telescopes, then, the relevant parameter space is well defined. We illustrate this parameter space in Fig.~\ref{fig:ps}.
%%%%%%%%%%%%%%%%%%%%%%%%%
%%%%%%%%

Eq.~(\ref{eq:L1}) is subject to various experimental constraints. In App.~\ref{app:expcst} we review the most relevant processes, summarized as follows:
\begin{itemize}
\item If $\phi$ is lighter than about 2~MeV, then the non-observation of neutrinoless double-beta decay involving the emission of a light degree of freedom imply $\G\lesssim10^{-5}$. The number 2~MeV corresponds to the available phase space for the reaction $(A,Z)\to(A,Z+2)+2e^-+\phi$. This lower limit on $m_\phi$ is comparable to the constraint due to the number of relativistic degrees of freedom during big-bang nucleosynthesis. 
\item Leptonic decays of $\pi$ and $K$ mesons imply $\G\lesssim10^{-2}$, irrespective of $m_\phi$ as long as $m_\phi\ll m_\pi\approx139.6$~MeV and $m_K\approx494$~MeV, respectively.
\end{itemize}

In the rest of this section we propose one example for a model that generates Eq.~(\ref{eq:L1}) with the small explicit breaking of U(1)$_L$ discussed above. We note that, in most respects, the details of this example are not crucial for the phenomenology at neutrino telescopes. However, within our specific model, additional constraints (arising from the mixing of heavy Dirac sterile neutrinos with the active neutrinos) imply that $\G\lesssim10^{-3}$. We provide further details on these additional constraints below in this section and in App.~\ref{app:expcst}.

As an example for generating Eq.~(\ref{eq:L1}), consider the potential
\be\label{eq:Luv0}V_{UV}&=&\left\{M\psi\psi^c+y'\Phi\psi^c\psi^c+y(HL)\psi+Y_lH^\dag Le^c+cc\right\}+m_\phi^2|\Phi|^2+\lambda_\phi|\Phi|^4+V_{\rm U(1)_{\not{L}}}\ee
with $m_\phi^2>0$. 
Here $\psi,\psi^c$ are Weyl spinor SM singlets, corresponding to heavy Dirac sterile neutrinos. We assume three generations of $\psi,\psi^c$. 
We assume that the mass $M$ is large compared to the energy scales of the problem, $M\gg m_\phi$. The basic setup corresponds to the inverse seesaw model~\cite{PhysRevLett.56.561,PhysRevD.34.1642,BernabŽu1987303} (see Ref.~\cite{Weiland:2013wha} for an overview) where the lepton number violating spurion $\mu$ is promoted to a field.

Setting $V_{\rm U(1)_{\not{L}}}=0$, the potential in Eq.~(\ref{eq:Luv0}) conserves lepton number and no neutrino mass is generated.
The effects we are interested in come from small lepton number violation encoded in $V_{\rm U(1)_{\not{L}}}$.  
To be concrete, we introduce a small tadpole for $\Phi$,
\be\label{eq:tphi} V_{\rm U(1)_{\not{L}}}=-t_\phi\Phi+cc.\ee
We assume $t_\phi\ll m_\phi^3$. This causes $\Phi$ to develop a VEV
\footnote{Instead of Eq.~(\ref{eq:tphi}) we could also break U(1)$_L$, for example, by introducing the terms $m^2\Phi^2+\kappa\Phi^3$ in the Lagrangian. The leading effect would be just to re-introduce the tadpole radiatively, $t_\phi\sim(\kappa m^2)/(4\pi)^2\log(\Lambda/m_\phi)$, where $\Lambda$ is some effective cut-off for the theory. For $\kappa\sim m\sim m_\phi$, with $\Lambda$ not too high above the weak scale, the resulting value of $t_\phi$ is consistent with the parameter space of interest to us here. The main difference between this possibility and the one suggested in Eq.~(\ref{eq:tphi}) is that now, the scalar and pseudo-scalar states in $\phi$ would be split in mass. This would modify our analysis in a straightforward manner, leaving our basic results unchanged.}
,
\be\langle \Phi\rangle&\equiv&\mu=\frac{t^*_\phi}{m_\phi^2}\ll m_\phi.\ee
For small $t_\phi$, the scalar and pseudo-scalar excitations in $\phi$ remain approximately degenerate with common mass $m_\phi$, and so we continue to treat $\phi$ as a single complex scalar state (we comment on the small breaking of scalar--psuedo-scalar mass degeneracy in App.~\ref{app:xs}).  
Integrating out $\psi$ and $\psi^c$ gives, to leading order in $M^{-1}$,
\be\label{eq:effG}\mathcal{L}_{\rm eff}&=&-(\mu+\phi)\,(HL)^T\,y^T\left(M^{-1}\right)^Ty'M^{-1}y\,(HL)+cc\\
&+&(HL)^\dag\,y^\dag\left(M^{-1}\right)^\dag M^{-1}\,y\,\bar{\sigma}^\mu i\partial_\mu\,(HL).\no\ee
This reproduces Eq.~(\ref{eq:L1}) together with additional terms (non-canonical neutrino kinetic term) that lead to additional, model-dependent constraints.

The tadpole $t_\phi$ could come, for example, from non-renormalizable operators generated at a high scale. To get an idea for the relevant scales, $t_\phi\sim(m^4/M_{pl})$, where $m\sim100$~GeV is at the weak scale and $M_{pl}\sim10^{19}$~GeV, would lead to\footnote{In this case it remains to be explained why quantum effects do not produce $t_\phi\sim M_{pl}^3$. One possible solution could be supersymmetry. We thank Takemichi Okui for a critical discussion regarding this point.} $t_\phi\sim10^{-2}$~MeV$^3$ and so $\mu\sim100$~eV for $m_\phi\sim10$~MeV, in accordance with Eq.~(\ref{eq:mureq}) and in the ballpark of the parameter space that can be probed by IceCube. 
Alternatively, U(1)$_L$ breaking could be spontaneous but occur in a slightly more complex scalar sector coupled to $\Phi$, in which case there will be some extra light Goldstone or pseudo-Goldstone boson states below $m_\phi$. In general, the smallness of $t_\phi$ (compared to the weak scale and to the mass $m_\phi$) is technically natural~\cite{'tHooft:1979bh} and there are many ways to produce this level of U(1)$_L$ breaking, the details of which are not important for the current paper. 

We note that constraints from $\mu\to e\gamma$ require that the parameters $M,\;y,\;y',\;Y_l$ of Eq.~(\ref{eq:Luv0}) exhibit nontrivial flavor structure. We do not analyze the model-building implications in detail. A consistent possibility is to assume that lepton flavor violation is contained only in the coupling $y'$, parametrizing the interactions of $\Phi$ with the fermion sector, while $M,y$ and $Y_l$ are taken to be lepton flavor universal. The $\Phi$ sector is then responsible for both the lepton number and the lepton flavor violation required by the observed neutrino masses and mixing.

In addition, the non-canonical neutrino kinetic terms in Eq.~(\ref{eq:effG}) lead to non-unitarity of the three-by-three Pontecorvo-Maki-Nakagawa-Sakata (PMNS) neutrino mixing matrix. This non-unitarity is severely constrained by precision flavor and electroweak laboratory experiments. We discuss these constraints in App.~\ref{app:expcst}, finding that they require that $\G\lesssim10^{-3}$ if we impose that the coupling $y'$ is perturbative, $y'\lesssim1$.

Lastly, we should emphasize again that the requirements in Eqs.~(\ref{eq:Gbig}) and~(\ref{eq:mureq}) -- the constraint on the scale of new physics $\Lambda$, and the requirement of explicit U(1)$_L$ violation that guided our discussion above -- are motivated by our focus on the phenomenology at high energy neutrino telescopes. From the theory point of view, alternative possibilities include spontaneous breaking of U(1)$_L$ (with interesting phenomenology in other settings, e.g.~\cite{Chacko:2003dt,Hall:2004yg,Davoudiasl:2005ks,Goldberg:2005yw,Baker:2006gm,Friedland:2007vv}), and different values of the scalar mass $m_\phi$ that, once we assume explicit U(1)$_L$ violation, is a free parameter of the model. In addition, it is also possible to imagine a scenario where the lepton number violating parameter $\mu$ would carry its own flavor structure, unrelated to the interactions of the scalar $\phi$, that would then have no obvious relation to neutrino mass generation. In this paper, however, we concentrate on the simple set-up given by Eqs.~(\ref{eq:L1}-\ref{eq:mu}). In the next section we work out in detail the implications at neutrino telescopes.

%%%%%%%%%%%%%%%%%%%%%%%%%%%%%%%%%%%%%
\section{Effects in neutrino telescopes}\label{sec:nutele}

\subsection{Boltzmann equation and optical depth}\label{ssec:bol}

To see when neutrino-neutrino scattering could affect the fluxes in neutrino telescopes, we need to evaluate the cosmological evolution of the neutrino flux. This is done using the Boltzmann equation.  
The Boltzmann equation for the comoving high energy neutrino density $N_i[q,\eta]$, assuming proper relic densities $n_i$ of non-relativistic target neutrinos, comoving energy $q=(1+z)\epsilon$ and conformal time $d\eta=a^{-1}c\,dt$ where $a=(1+z)^{-1}$ is the scale factor, is
\be\partial_\eta N_i\left[q;\eta\right]&=&
aQ_i\left[q,\eta\right]\\
&-&acN_i[q;\eta]\Big\{\sum_jn_i(\eta)\sigma_{ii\to jj}[2m_{\nu_i}q]
+\sum_{j\neq i}n_j(\eta)\sigma_{ij\to ij}[2m_{\nu_j}q]\Big\}\no\\
&+&ac\int_0^1 \frac{dx}{x}\Big\{\sum_j\frac{d\sigma_{jj\to ii}\left[\frac{2m_{\nu_j}q}{x},x\right]}{dx}N_j\left[\frac{q}{x},\eta\right]n_j(\eta)\no\\
&+&\sum_{j\neq i}\frac{d\sigma_{ij\to ij}\left[\frac{2m_{\nu_j}q}{x},x\right]}{dx}N_i\left[\frac{q}{x},\eta\right]n_j(\eta)
+\sum_{j\neq i}\frac{d\sigma_{ji\to ji}\left[\frac{2m_{\nu_i}q}{x},x\right]}{dx}N_j\left[\frac{q}{x},\eta\right]n_i(\eta)\Big\}.\no\ee
We define $Q_i[q,\eta]$ as the comoving injection rate density. In what follows we move freely between redshift $z$ and conformal time $\eta$. We use $n_i=112\,(1+z)^3$~cm$^{-3}$, assuming an equal mix of C$\nu$B neutrinos and antineutrinos.   
In App.~\ref{app:xs} we collect formulae for the total and differential scattering cross sections, $\sigma[\hat s]$ and $d\sigma[\hat s,x]/dx$, respectively, where $\sqrt{\hat s}$ is the center of mass energy and $x$ is the inelasticity.  

It is useful to define the optical depth, $\tau$, for a neutrino with observed energy $\epsilon$ that was emitted at a time $\eta$, scattering through some cross section $\sigma[\hat s]$, as
\be\tau[q;\eta]&=&\int_{\eta}^{\eta_0} d\eta'a'cn(\eta')\sigma[2m_\nu q'],\;\;\;\;\;\;\;\dot\tau=\partial_\eta\tau=-acn(\eta)\sigma[2m_\nu q].\ee
In the integrand, $q'=(1+z')\epsilon$. Note that $\eta_0\approx1.4\times10^4$~Gpc. The redshifts of interest are $z\lesssim 5$ or so, when astrophysical sources are likely to be active. An optical depth of order unity implies significant scattering effects in the neutrino flux.

The process $ij\to ij$ proceeds via t-channel scattering, and is a smooth function of the center of mass energy $\sqrt{\hat s}$ and inelasticity $x$. For energy-independent scattering, one has
\be\tau&\approx&\frac{c\sigma n(0)}{H_0}\frac{2}{3\Omega_m}\left(\sqrt{\Omega_\Lambda+\Omega_m\left(1+z\right)^3}-1\right)\approx\left(\frac{\sigma}{10^{-31}~\rm{cm^2}}\right)\left(\frac{n(z=0)}{100~\rm{cm^{-3}}}\right)\left(\frac{1+z}{3}\right)^{\frac{3}{2}}.\ee
The cross section $\sigma_{ij\to ij}$ is maximized around $\hat s\approx1.6m_\phi^2$, with peak cross section $\sigma_{ij\to ij}({\rm peak})\approx\left|\G_{i}\right|^2\left|\G_{j}\right|^2/(260\pi m_\phi^2)\approx 5 \times 10^{-35}\left(\mathcal{G}/10^{-2}\right)^4\left(m_\phi/10~{\rm MeV}\right)^{-2}$~cm$^2$. This leads to an optical depth
\be\label{eq:tijij}\tau_{ij\to ij}\sim5\times10^{-8}\left(\frac{\G}{10^{-3}}\right)^4\left(\frac{m_\phi}{10~{\rm MeV}}\right)^{-2}\left(\frac{1+z}{3}\right)^{\frac{3}{2}}.\ee
As discussed at the end of Sec.~\ref{sec:model}, laboratory constraints require that $m_\phi\gtrsim2$~MeV and $\G\lesssim10^{-2}$. Plugging these values into Eq.~(\ref{eq:tijij}) we find that the t-channel optical depth in our model cannot be significant, $\tau_{ij\to ij}\ll1$. 

The second class of processes we have involve resonant s-channel scalar exchange, occurring in $ii\to jj$ transitions. 
Close to the resonance we can write the cross section as
\be\label{eq:deltaapp} \sigma_{ii\to jj}&\approx&\frac{|\G_{i}|^2|\G_{j}|^2}{64\pi\Gamma_\phi^2}\frac{m_\phi^2\Gamma_\phi^2/(4m_{\nu_i}^2)}{\left(\epsilon-\frac{m_\phi^2}{2m_{\nu_i}}\right)^2+m_\phi^2\Gamma_\phi^2/(4m_{\nu_i}^2)}\\
&\approx&\frac{\pi}{2}\frac{|\G_{i}|^2|\G_{j}|^2}{\sum_k|\G_{k}|^2}\frac{1}{m_\phi^2}\delta\left(1-\frac{m_\phi^2}{2m_{\nu_i}\epsilon}\right),\no\ee
where in the second line we assumed that $\phi$ decays predominantly back into neutrinos, using the expression for $\Gamma_\phi$ given in Eq.~(\ref{eq:Gs}). 
Using the delta function approximation gives
\be\label{eq:tiijj}\tau_{ii\to jj}[q;\eta]&\approx&\frac{\pi}{2}\frac{|\G_{i}|^2|\G_{j}|^2}{\sum_k|\G_{k}|^2}\frac{c}{m_\phi^2}\int_\eta^{\eta_0} d\eta'a'n_i(\eta')\delta\left(1-\frac{m_\phi^2}{2m_{\nu_i}q'}\right)\\
%
%&=&\frac{\pi}{2}\frac{|\G_{i}|^2|\G_{j}|^2}{\sum_k|\G_{k}|^2}\frac{c}{m_\phi^2}\,\frac{n_i(0)\,\left(\frac{m_\phi^2}{2m_{\nu_i}aq}\right)^3}{H_0\,\sqrt{\Omega_\Lambda+\Omega_m\left(\frac{m_\phi^2}{2m_{\nu_i}aq}\right)^3}}\,\theta\left(\frac{m_\phi^2}{2m_{\nu_i}}-aq\right)\,\theta\left(q-\frac{m_\phi^2}{2m_{\nu_i}}\right)\no\\
%
&\sim&8\,\left(\frac{\G}{10^{-3}}\right)^2\left(\frac{m_\phi}{10~{\rm MeV}}\right)^{-2}\,\left(\frac{m_\phi^2}{2m_{\nu_i}\epsilon}\right)^3\,\theta\left(\frac{m_\phi^2}{2m_{\nu_i}}-\epsilon\right)\,\theta\left((1+z)\epsilon-\frac{m_\phi^2}{2m_{\nu_i}}\right).\no\ee

For our model, therefore, because of the constraint $\mathcal G\lesssim10^{-2}$, resonant s-channel scattering could become significant while the non-resonant t-channel processes are unimportant. 
Using the delta function approximation for the resonant part of the cross section and neglecting the non-resonant processes, we can rewrite the optical depth and the Boltzmann equation as
\be\label{eq:taudlta}\dot\tau_{ii\to jj}[q,\eta]&=&-\frac{\pi acn_i}{2m_\phi^2}\frac{|\G_{i}|^2|\G_{j}|^2}{\sum_k|\G_{k}|^2}\delta\left(1-\frac{m_\phi^2}{2m_{\nu_i}q}\right)\equiv-\delta\left(1-\frac{m_\phi^2}{2m_{\nu_i}q}\right)\,\dot{\bar\tau}_{ii\to jj}(\eta),\\
\tau_{ii\to jj}[q,\eta]&=&
\theta\left(q-\frac{m_\phi^2}{2m_{\nu_i}}\right)\theta\left(\frac{m_\phi^2}{2m_{\nu_i}}-aq\right)\frac{\dot{\bar\tau}_{ii\to jj}(\eta')}{H(\eta')}\Big|_{a'=\frac{2m_{\nu_i}aq}{m_\phi^2}},\ee
and
\be\label{eq:BEres}
\left(\partial_\eta -\sum_j\dot\tau_{ii\to jj}[q,\eta]\right)N_i\left[q;\eta\right]
&=&aQ_i\left[q,\eta\right]
+2\sum_j\theta\left(\frac{m_\phi^2}{2m_{\nu_j}}-q\right)\dot{\bar\tau}_{jj\to ii}(\eta)N_j\left[\frac{m_\phi^2}{2m_{\nu_j}},\eta\right].\no\\
\ee
We solve Eq.~(\ref{eq:BEres}) numerically, beginning at high redshift with vanishing high energy neutrino flux. 

In the calculation above we assumed that $\phi$ decays predominantly into neutrinos, using Eq.~(\ref{eq:Gs}). With this assumption, neutrino scattering leads to a deficit in the observed neutrino flux close to the resonance energy, and to a pile-up at lower energy due to regeneration.  This assumption, however, is not necessarily true. For instance, if lepton number is broken spontaneously by some scalar sector that couples to $\Phi$, then $\phi$ may decay mostly to Goldstone bosons in that sector. In this case the scattering formalism remains the same, but the regeneration term ($\dot{\bar{\tau}}$ term on the RHS) in Eq.~(\ref{eq:BEres}) is eliminated\footnote{We comment that, without regeneration, Eq.~(\ref{eq:BEres}) is easily solved analytically.}. In addition, the scattering cross section and rate in Eqs.~(\ref{eq:deltaapp})-(\ref{eq:taudlta}) should be multiplied by a factor of $BR(\phi\to\nu\nu)$, implying that a larger value of $\G$ would be required to achieve a given optical depth.

\subsection{Features in the neutrino flux and flavor composition}

We are now in position to analyze the neutrino flux and flavor composition at neutrino telescopes. 
To this end, we need to make some assumptions regarding the astrophysical neutrino sources. 
In this work, the astrophysical neutrino source is assumed to have constant power per log energy bin, $\epsilon_\nu^2Q\propto$~const. Assuming that cosmic rays are accelerated by stochastic processes \cite{Fermi:1949ee}, a simple power-law spectrum is naturally expected, especially if the neutrinos are produced by $pp$ interactions. This is the case, for example, in cosmic ray reservoirs such as galaxies and galaxy clusters~\cite{Murase:2013rfa,Katz:2013ooa}.  
Note that high energy neutrinos may also be produced via $p\gamma$ interactions, in which case neutrino spectra depend on details of target photon fields and may not be a simple power law (e.g., Refs.~\cite{Stecker:2013fxa,Murase:2013ffa,Winter:2013cla}).  For example, in the classical scenario of gamma-ray bursts (GRBs), $\epsilon_\nu^2Q\propto$~const was expected from $\epsilon_\nu\sim100$~TeV to $\epsilon_\nu\sim100$~PeV~\cite{Waxman:1997ti}, with breaks at lower and higher energy. 
Clearly, a better understanding of the astrophysical neutrino source would be required in order for the scattering effects described here to be identified. Some analysis of the joint effect of neutrino scattering together with deviations from power-law form in the input astrophysical flux can be found in Refs.~\cite{Ng:2014pca,Ioka:2014kca}.

Considering the flavor composition, we assume $\nu_e:\bar{\nu}_e:\nu_\mu:\bar{\nu}_\mu:\nu_\tau:\bar{\nu}_\tau\approx1:1:2:2:0:0$ at the source. This assumption is consistent with the expectation if neutrino production occurs through $pp$ collisions or $p\gamma$ interactions in the multipion production region.  
For $p\gamma$ interactions when $\Delta$ resonance and direct production dominate, one expects $\nu_e:\bar{\nu}_e:\nu_\mu:\bar{\nu}_\mu:\nu_\tau:\bar{\nu}_\tau\approx1:0:1:1:0:0$ at the source. Assuming that the C$\nu$B contains an equal mix of neutrinos and antineutrinos, the deficit of antineutrinos in $p\gamma$ production does not affect the scattering rate in our model. 
In either of the $pp$ or $p\gamma$ scenario, $\nu_e+\bar{\nu}_e:\nu_\mu+\bar{\nu}_\mu:\nu_\tau+\bar{\nu}_\tau\approx1:2:0$, leading to $\nu_e+\bar{\nu}_e:\nu_\mu+\bar{\nu}_\mu:\nu_\tau+\bar{\nu}_\tau\approx1:1:1$ at the Earth~\cite{Learned:1994wg,Beacom:2003nh} in the absence of new interactions.  
We comment that strong cooling of mesons and muons at the source could in principle affect the flavor ratios (e.g., Refs.~\cite{Rachen:1998fd,Kashti:2005qa,Rodejohann:2006qq}), but this effect should be accompanied by a spectral suppression that would be identifiable at IceCube. In addition, the flavor composition at earth varies by a few percent (e.g., Refs.~\cite{Kashti:2005qa,Winter:2006ce,Rodejohann:2006qq,Blum:2007ie}) when neutrino mixing parameters are varied within their current experimental range~\cite{Capozzi:2013csa}. 

It has been known that most astrophysical sources evolve with redshift.  
For concreteness, we adopt a redshift evolution using the parametrization of~\cite{Yuksel:2006qb} for the GRB rate, 
\begin{equation}
 \frac{Q[\epsilon,z]}{Q[\epsilon,z=0]}\propto
\left\{\begin{array}{ll}
{(1+z)}^{4.8} 
& \mbox{(for $z\leq1$)},
\\
{(1+z)}^{1.4}
& \mbox{(for $1<z<4.5$)},
\\
{(1+z)}^{-5.6} 
& \mbox{(for $4.5\leq z$)}.
\end{array} \right.
\end{equation}
Applications to other evolution models are straightforward. We comment that using the star formation rate, taken for instance from~\cite{Hopkins:2006bw}, gives comparable results, though slightly less favorable for detection.

\begin{figure}[!h]\begin{center}
\includegraphics[width=0.6\textwidth]{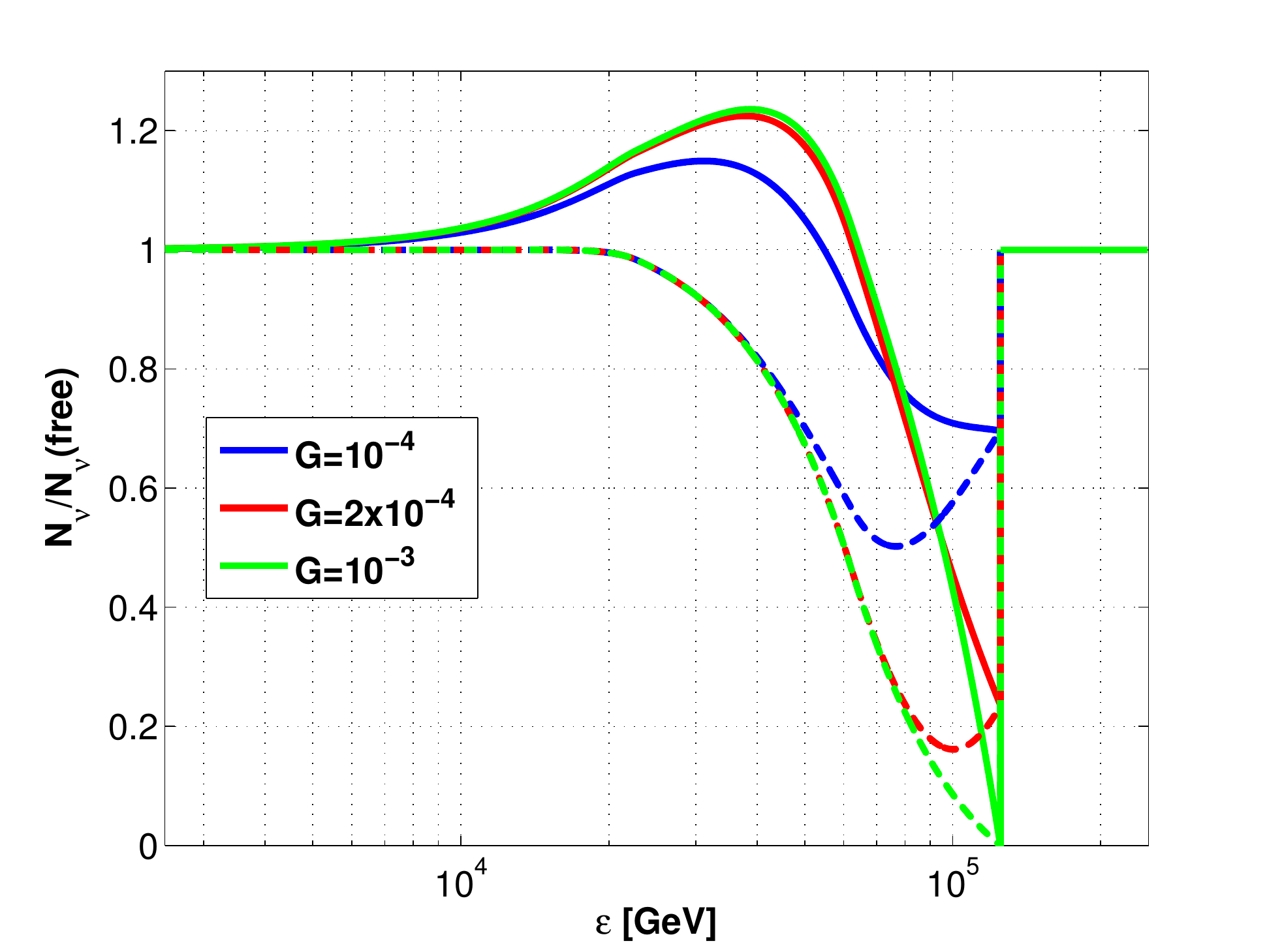}\end{center}
\caption{Ratio of scattered to free neutrino flux as a function of energy when only a single flavor scatters using the narrow resonance delta function approximation for the cross section. Smooth (dashed) lines show the result with (without) regeneration. The value of the coupling is noted in the legend. The scalar mass is $m_\phi=5$~MeV and the neutrino mass is $\Sigma_i m_{\nu_i}=0.1$~eV. The astrophysical neutrino source is assumed to have a power low form in energy, $\epsilon^2Q\propto$~const, and follows the redshift evolution using the parametrization of~\cite{Yuksel:2006qb}. %~\cite{Hopkins:2006bw} .
}
\label{fig:D}
\end{figure}

Turning finally to the neutrino flux, consider first a single neutrino example omitting flavor effects. Examples are shown in Fig.~\ref{fig:D} for varying values of $\G$.  Smooth (dashed) lines show the result with (without) regeneration. The spectral dip is a result of resonant scattering where neutrinos with $\epsilon = m_\phi^2/2 m_\nu$ are scattered to lower energies. The rise at lower energies is a result of the higher energy neutrinos being scattered into the lower energy bins. The step-like drop at the dip is an artifact of our delta approximation, that is only expected to hold for small values of the coupling $\G$ or, more precisely, a narrow width for $\phi$. Similar calculations where done in Ref.~\cite{Ng:2014pca}, that assumed a Breit-Wigner form for the scattering cross section and did not use the delta approximation (and allowed large values of the coupling $\G$, that would be excluded by laboratory constraints for the particular model we analyze here). We find reasonable agreement with their results when considering models with a small coupling $\G\sim10^{-2}$. As we demonstrate in Sec.~\ref{ssec:detect}, we expect that the limited energy resolution of a neutrino telescope would wash out most of the error due to our delta approximation.  
%%%%%%%%%
%%%%%%%%

Next we work out the prediction of the model of Sec.~\ref{sec:model}, with $\G_{i}\propto m_{\nu_i}$. At this point we add flavor information, in order to translate the scattering -- best described in terms of neutrino mass eigenstates $\nu_1,\nu_2,\nu_3$ -- to the gauge eigenstates $\nu_e,\nu_\mu,\nu_\tau$ observed at the detector. Our treatment of neutrino flavor mixing during propagation is as follows. We assume that the cosmological sources produce a neutrino beam with well-defined flavor content in the gauge interaction basis, such that the flux of neutrino flavor $\alpha$ is $J^{\rm source}_{\nu_\alpha}$ for $\alpha=e,\mu,\tau$. For all scenarios of interest to us here, the mean free path for neutrino scattering is many orders of magnitude larger than the neutrino oscillation length, $l_{\rm osc}\ll l_{\rm mfp}$. This means that by the time that a neutrino beam of initial flavor $\alpha$ from a cosmological source first scatters, its flavor state would correspond to a statistical ensemble with probability $P_{i\alpha}=|U_{\alpha i}|^2$ to be found as mass-eigenstate $i$. The flux of neutrinos of mass eigenstate $i$, given at intermediate distance $l_{\rm osc}\ll l\ll l_{\rm mfp}$, is then $J^{\rm source}_{\nu_i}=P_{i\alpha}J^{\rm source}_{\nu_\alpha}$. We now subject the input flux  $J^{\rm source}_{\nu_i}$ to scattering and cosmological redshift evolution as described in Sec.~\ref{ssec:bol}. The propagated mass-eigenstate flux at the detector, $J^{\rm detector}_{\nu_i}$, is then transformed to the corresponding flavor flux via $J^{\rm detector}_{\nu_\alpha}=P_{i\alpha}J^{\rm detector}_{\nu_i}$.

\begin{figure}[!t]\begin{center}
\includegraphics[width=0.475\textwidth]{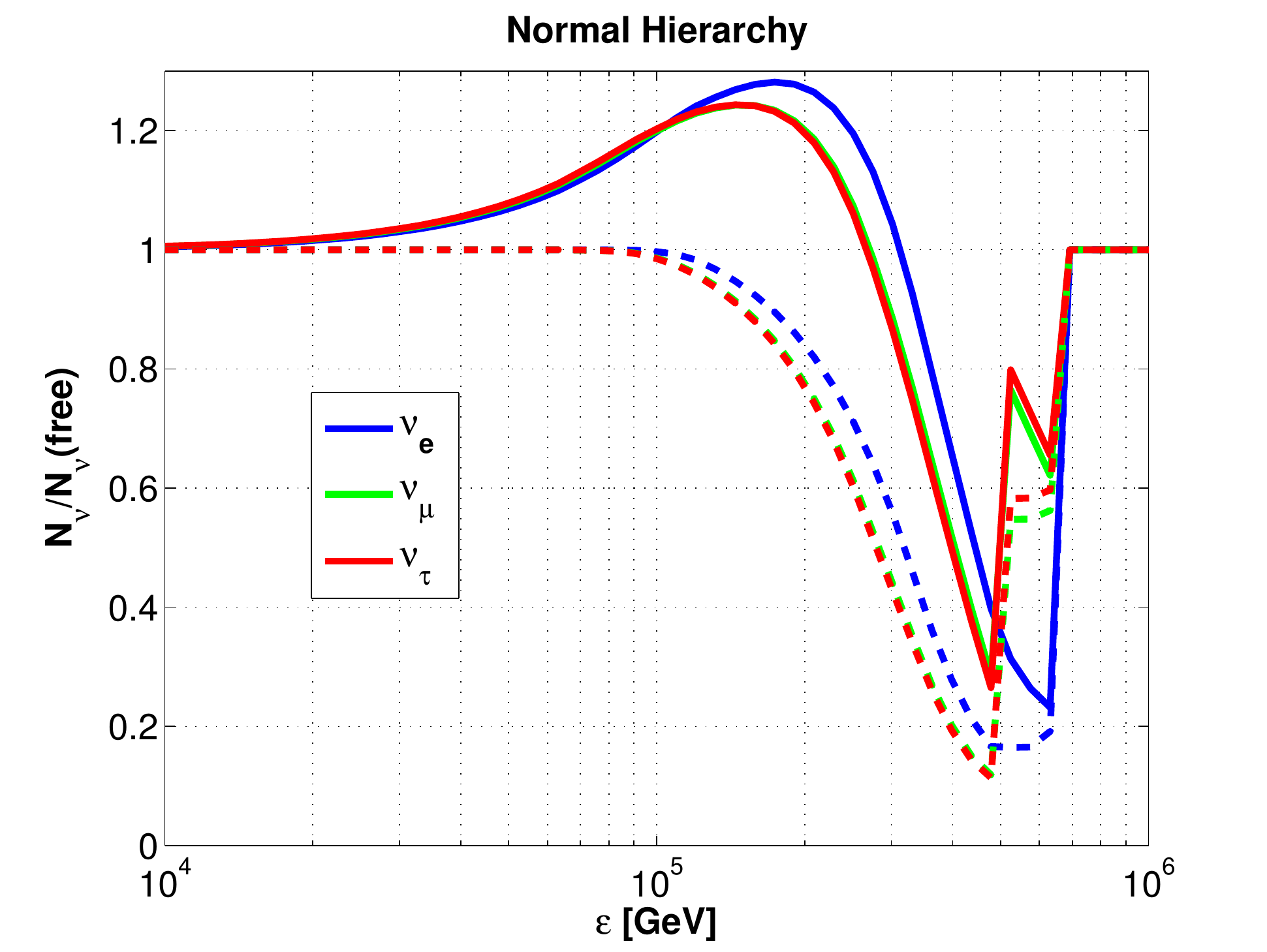}\quad
\includegraphics[width=0.475\textwidth]{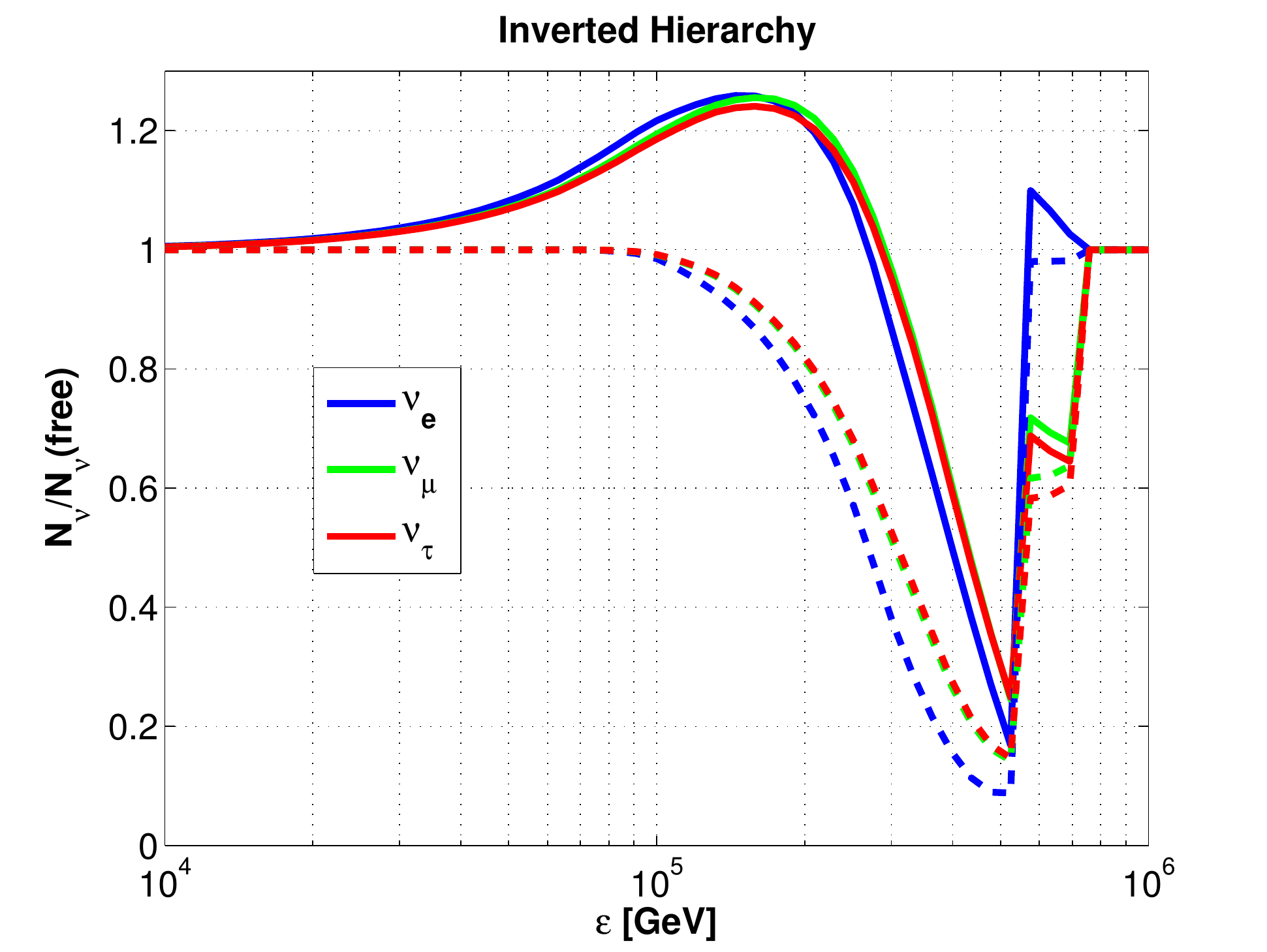} \\
\includegraphics[width=0.475\textwidth]{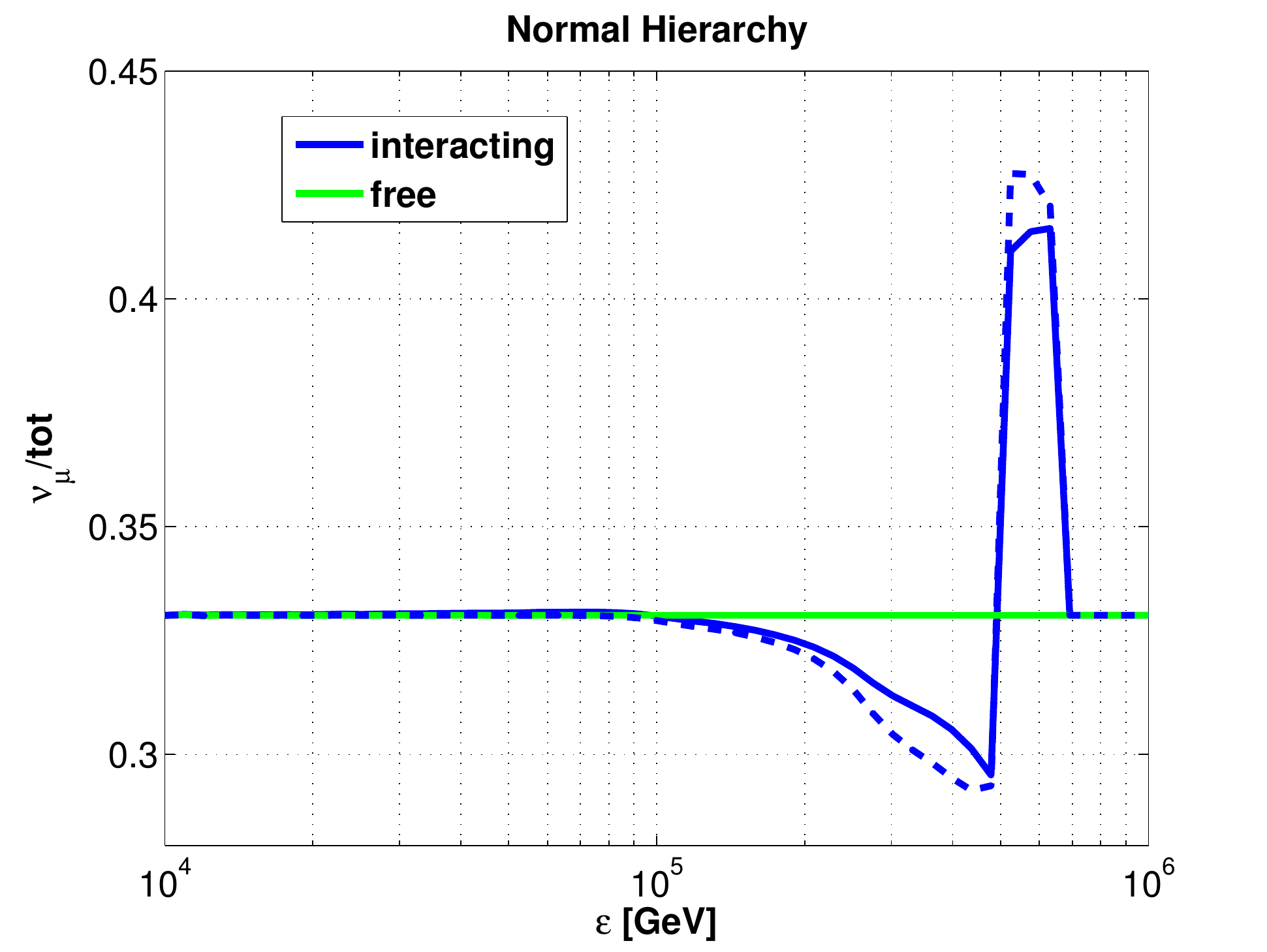}\quad
\includegraphics[width=0.475\textwidth]{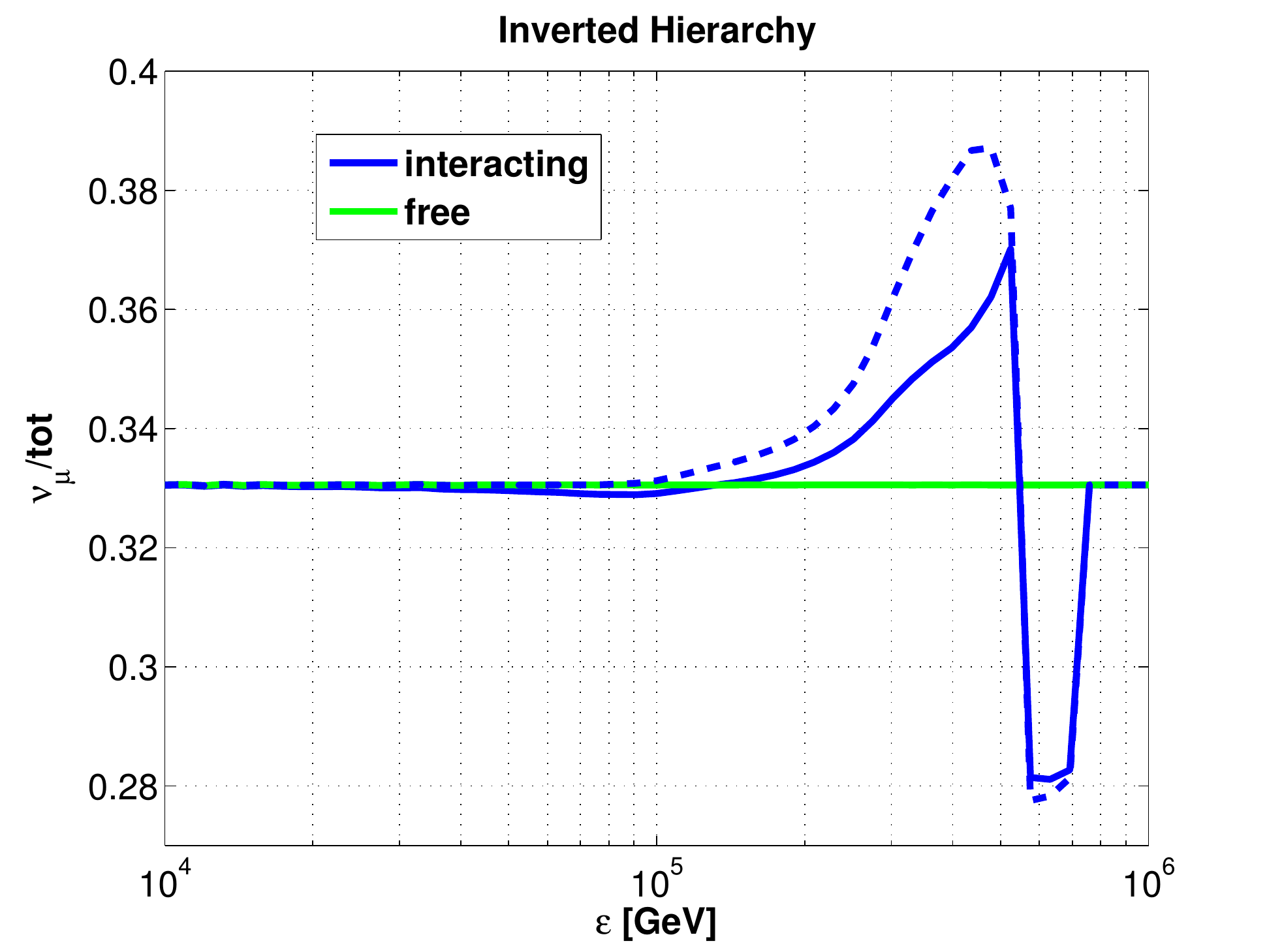}
 \end{center}
\caption{Top panels: neutrino flavor fluxes at Earth normalized to the case of free propagation. Bottom panels: muon neutrino to total neutrino flux ratio. Smooth (dashed) lines show the case with (without) regeneration. The scalar mass is $m_\phi=9$~MeV, the sum of neutrino masses is $\sum_im_{\nu_i}=0.2$~eV, and $\G=1.3\times10^{-3}$. The astrophysical neutrino source is the same as in Fig.~\ref{fig:D}, assuming flavor ratio corresponding to pion decay.}
\label{fig:Dflav}
\end{figure}

\begin{figure}[!t]\begin{center}
\includegraphics[width=0.475\textwidth]{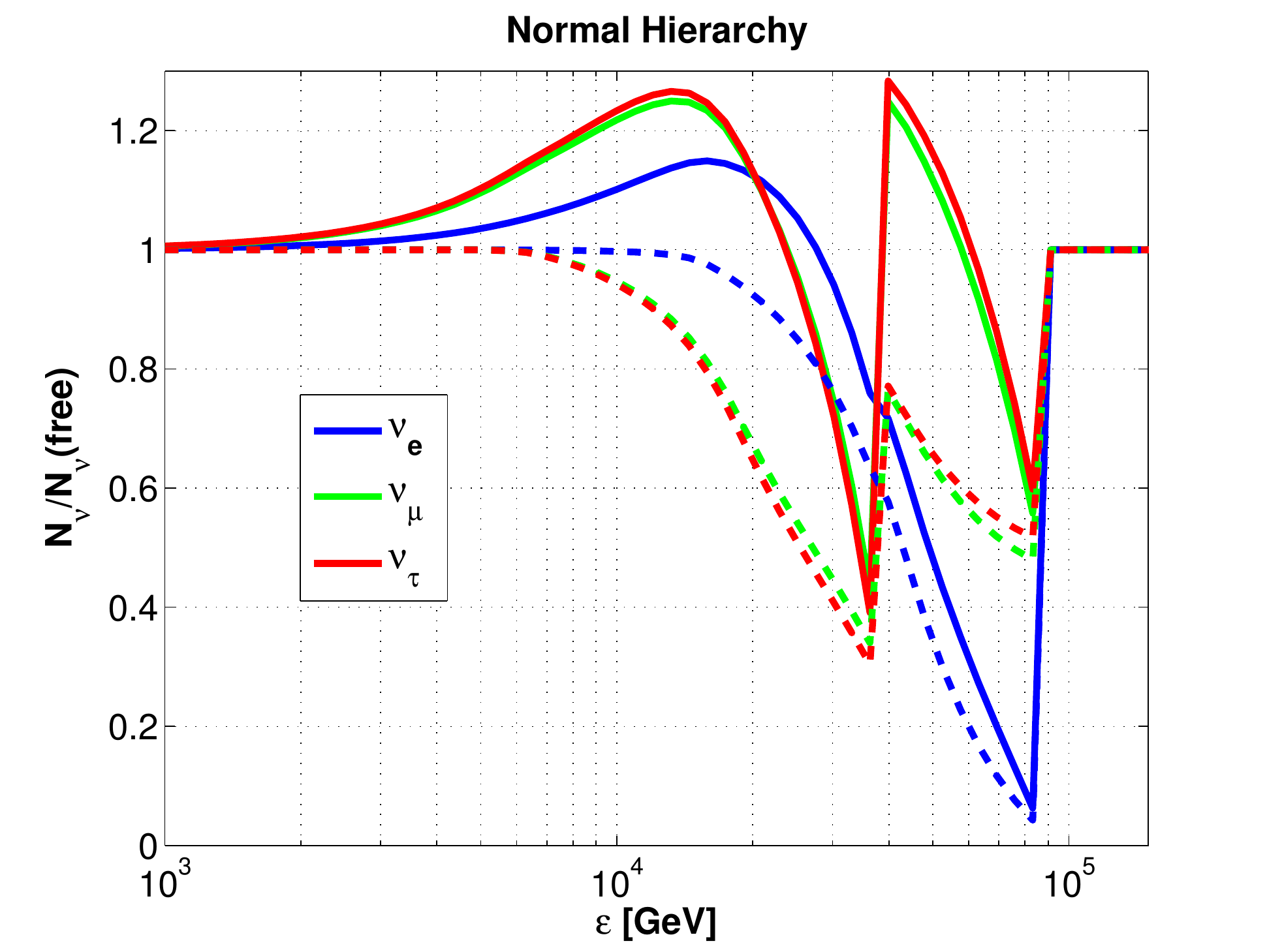}\quad
\includegraphics[width=0.475\textwidth]{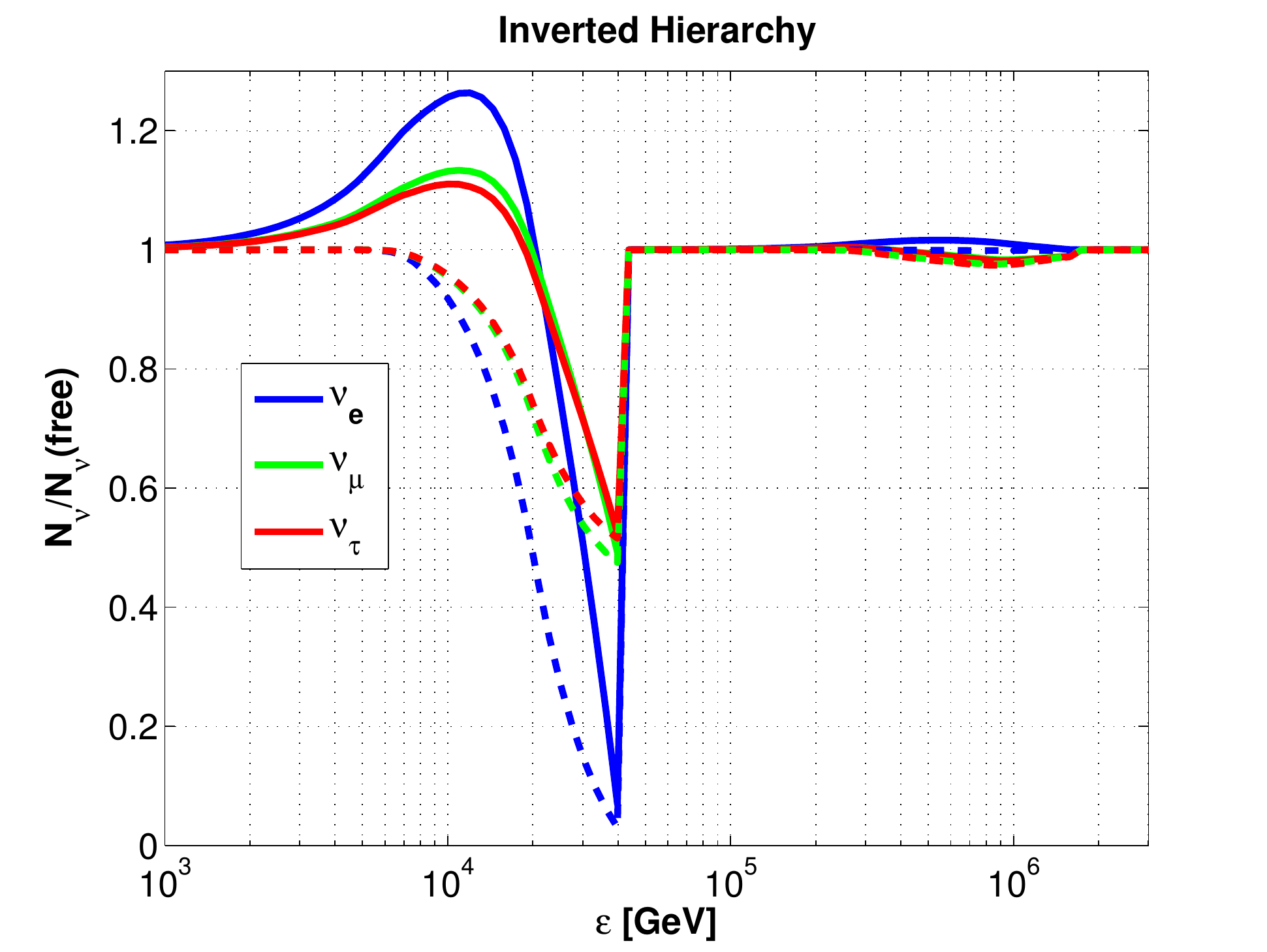} \\
\includegraphics[width=0.475\textwidth]{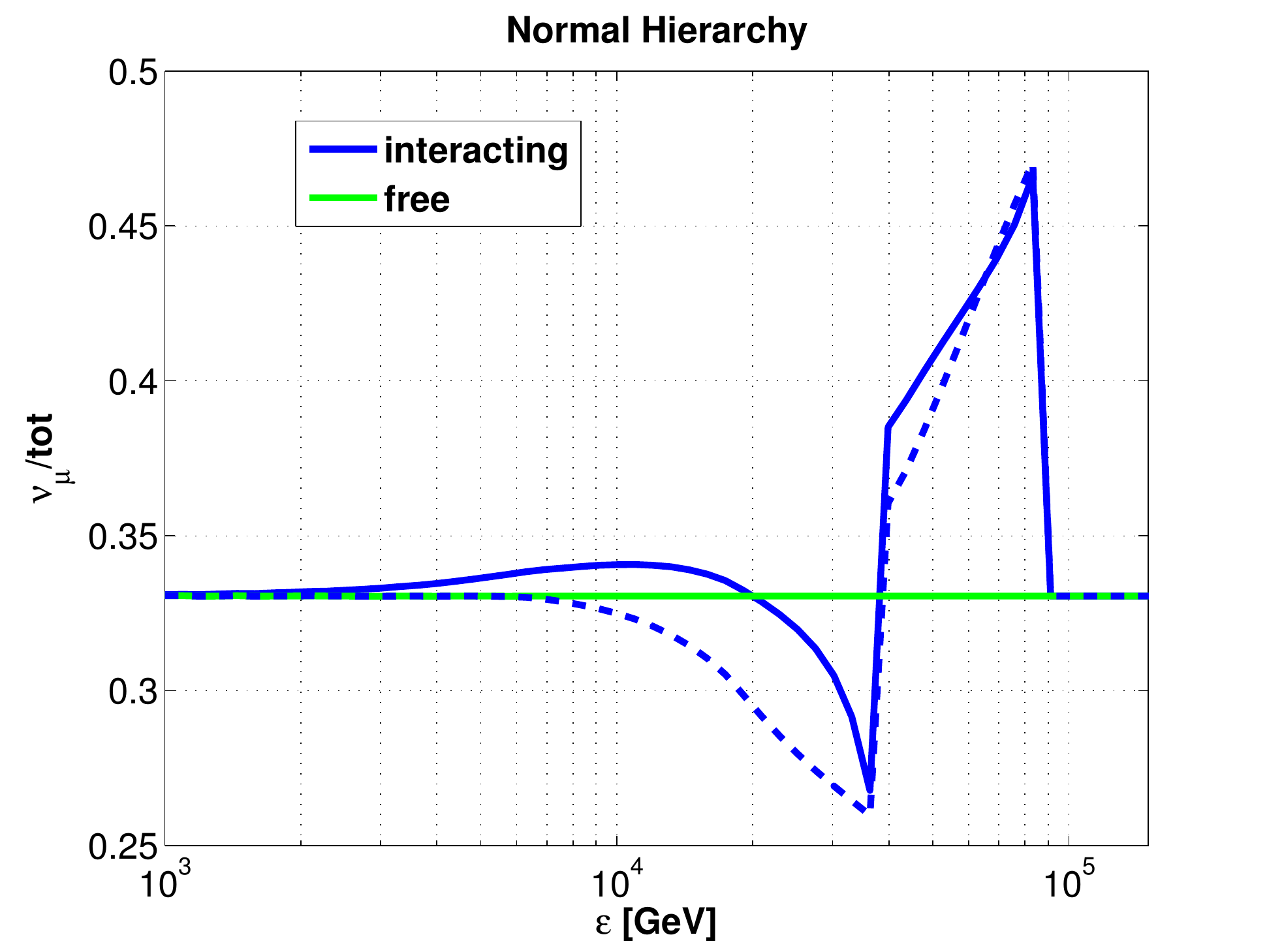}\quad
\includegraphics[width=0.475\textwidth]{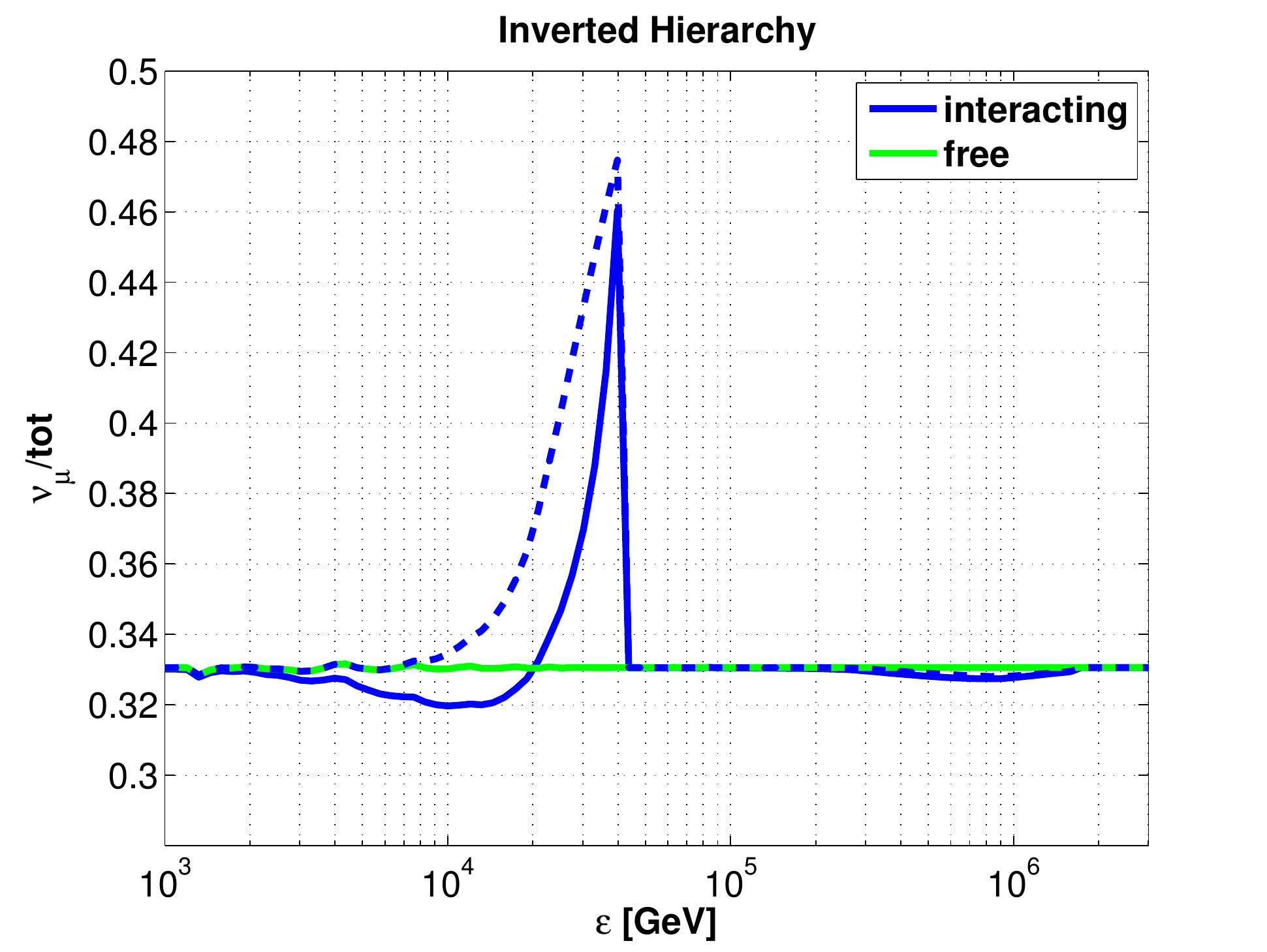}
 \end{center}
\caption{As in Fig.~\ref{fig:Dflav}, but using scalar mass $m_\phi=2$~MeV, sum of neutrino masses $\sum_im_{\nu_i}=0.1$~eV, and $\G=8\times10^{-4}$. }
\label{fig:Dflav2}
\end{figure}

In Fig.~\ref{fig:Dflav} we show the results for scalar mass $m_\phi=9$~MeV, sum of neutrino masses $\sum_im_{\nu_i}=0.2$~eV, and $\G=1.3\times10^{-3}$ corresponding to $\G_1\approx3.79\times10^{-4},\;\G_2\approx3.83\times10^{-4},\;\G_3\approx4.87\times10^{-4}$ in the case of normal neutrino mass hierarchy and $\G_1\approx3.36\times10^{-4},\;\G_2\approx4.55\times10^{-4},\;\G_3\approx4.58\times10^{-4}$ for inverted hierarchy. In Fig.~\ref{fig:Dflav2} we show the results for scalar mass $m_\phi=2$~MeV, sum of neutrino masses $\sum_im_{\nu_i}=0.1$~eV, and $\G=8\times10^{-4}$ corresponding to $\G_1\approx1.78\times10^{-4},\;\G_2\approx1.91\times10^{-4},\;\G_3\approx4.31\times10^{-4}$ in the case of normal hierarchy and $\G_1\approx0.09\times10^{-4},\;\G_2\approx3.92\times10^{-4},\;\G_3\approx3.98\times10^{-4}$ for inverted hierarchy. 
We use the neutrino oscillation parameters of~\cite{Capozzi:2013csa}.
%%%%%%%%%%%%%%%%%%%%%%%%%%%%
%%%%%%%%%%%%%%%%%%%%%%%%%%%%
%%%%%%%%%%%%%%%%%%%%%%%%%%%%
%%%%%%%%%%%%%%%%%%%%%%%%%%%%

A number of notable features are seen in Figs.~\ref{fig:Dflav} and~\ref{fig:Dflav2}.
For both normal and inverted mass hierarchies, the flux of muon and tau neutrinos relative to the flux they would have without scattering is equal to a few percent. This $\mu-\tau$ coincidence is a combined result of the approximate tri-bimaximal mixing of the PMNS matrix; the flavor pattern of the neutrinos emitted from the astrophysical sources, which we assumed to be pion decay; and our assumption that the scattering is proportional to powers of neutrino mass.

The flux of electron neutrinos deviates from that of muon and tau. 
The absorption dips track the resonance energy dictated by the neutrino mass eigenvalues, with lighter neutrinos producing a feature at higher energy and vice versa. In the case of the inverted hierarchy, the lightest neutrino mass eigenstate has a very small projection on $\nu_e$, and so the highest energy absorption dip is absent in $\nu_e$. In contrast, for normal hierarchy, the lightest mass eigenstate has a sizable projection on $\nu_e$, and the highest energy absorption dip is clearly seen in this flavor. At IceCube, this flavor effect is interesting as it causes a variation in the track to cascade event rate ratio.

\begin{figure}[!t]\begin{center}
\includegraphics[width=0.485\textwidth]{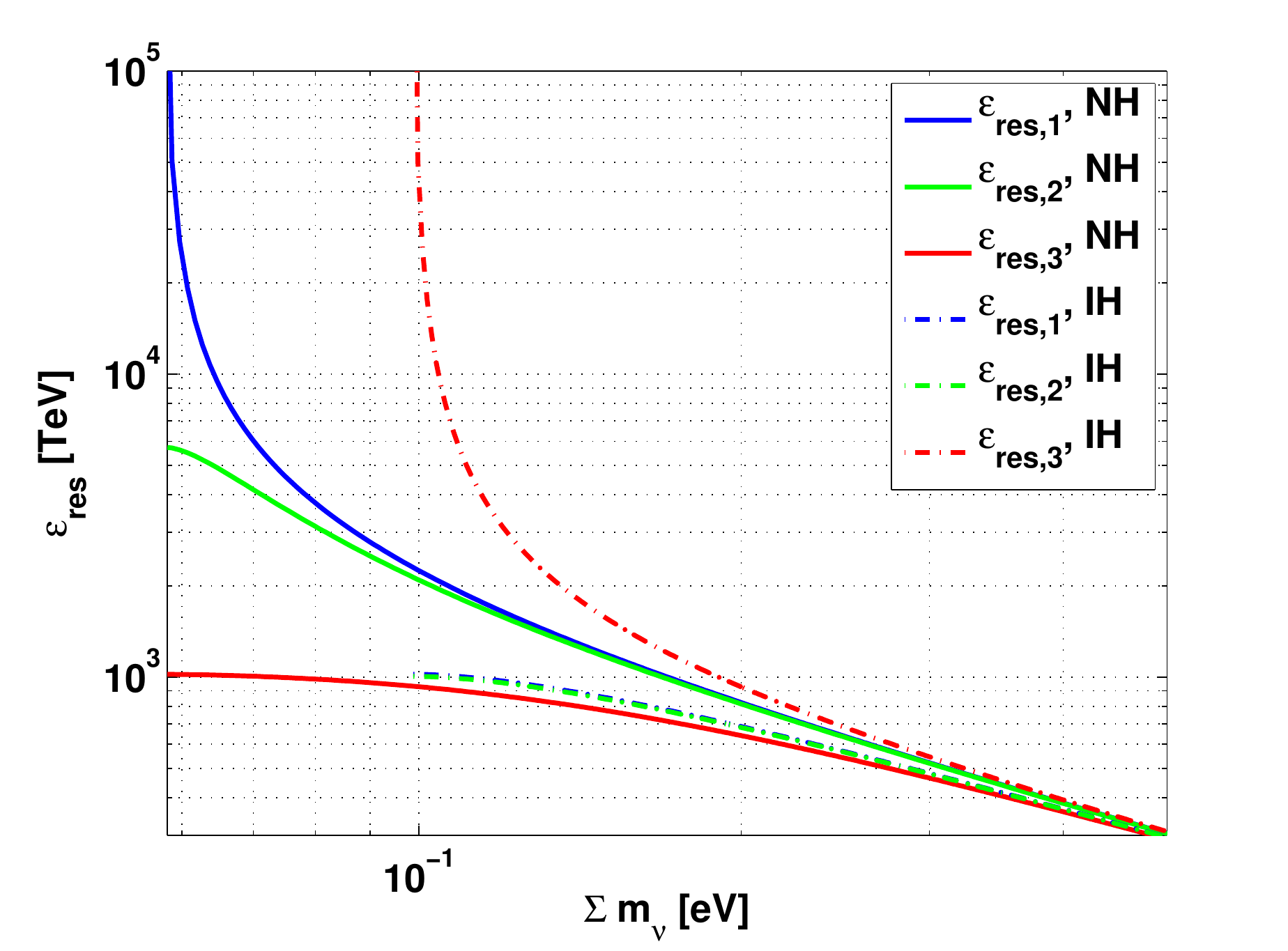}\quad
\includegraphics[width=0.485\textwidth]{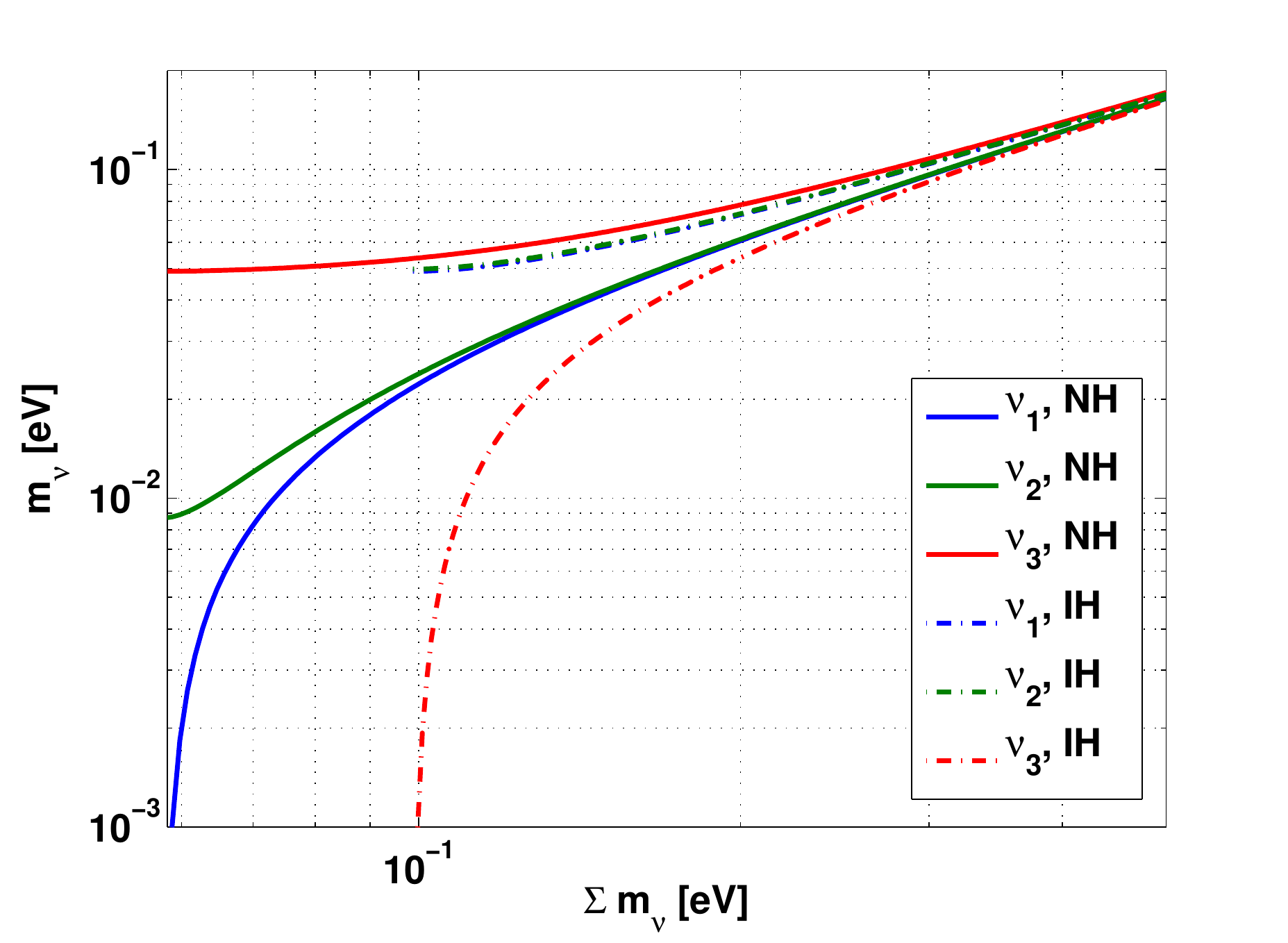} 
 \end{center}
\caption{Left: resonance energy as function of the sum of neutrino masses, for $m_\phi=10$~MeV. Right: neutrino mass eigenvalues. Smooth (dashed) lines refer to normal (inverted) neutrino mass hierarchy.}
\label{fig:numass}
\end{figure}

External information on the sum of neutrino masses from cosmology (see, e.g.~\cite{Hannestad:2014voa}) can provide consistency checks on the interpretation of a signal. On the left panel of Fig.~\ref{fig:numass} we illustrate this information by plotting the resonance energy as function of the sum of neutrino masses for fixed $m_\phi=10$~MeV. For convenience, on the right panel of Fig.~\ref{fig:numass} we plot the neutrino mass eigenvalues, using neutrino oscillation data from~\cite{Capozzi:2013csa}. Current cosmological data~\cite{Hinshaw:2012aka,Ade:2013zuv,Mantz:2014paa} suggests that $\sum_im_{\nu_i}<0.3$~eV. From Fig.~\ref{fig:numass}, this implies  $m_\nu\lesssim0.1$~eV for any mass eigenstate. Together with the constraint $m_\phi>2$~MeV, this already implies $\epsilon_{\rm res}\gtrsim20$~TeV.
%%%%%%%%%%%%%%%%%%%%%%%%%%%%
%%%%%%%%%%%%%%%%%%%%%%%%%%%%

\subsection{Detectability}\label{ssec:detect}

We now discuss observational prospects.  Our goal is to make a crude estimate of the opportunities and challenges for actual detection of anomalous neutrino interactions at IceCube, including issues such as backgrounds, energy resolution, and flavor identification. While a dedicated experimental analysis would be needed for any convincing detection, our analysis here is sufficient to get an estimate of the time scale before such detection could be expected, and of the model-building requirements that must be met if such detection is to become possible. Following Ref.~\cite{Laha:2013lka}, we take a simplified approach to calculate neutrino event rates in the IceCube detector.  As in Ref.~\cite{Aartsen:2013bka,Aartsen:2013jdh,Aartsen:2014gkd}, we mainly consider contained-vertex events, since energy resolution is better than for through-going muon events~\cite{Beacom:2004jb}.  We comment on the utility and limitations of through-going muon events at the end.   

Using a neutrino effective area $A_{\rm eff}(\epsilon_\nu)$ for each flavor~\cite{Aartsen:2013jdh}, event rates of neutrinos ranging from $\epsilon_{\nu1}$ to $\epsilon_{\nu2}$ are calculated by
\begin{equation}
{\mathcal N_{\nu}}^{[\epsilon_{\nu1},\epsilon_{\nu2}]} = \int_{\epsilon_{\nu1}}^{\epsilon_{\nu2}} d \epsilon_\nu \int d \Omega \,\,\, A_{\rm eff} (\epsilon_\nu) J_\nu(\epsilon_\nu).
\end{equation}
The above equation is applied to both shower-like and track-like events.  To compare with the data, it is useful to use deposited energy $E_{\rm dep}$, which is the energy neutrinos leave in the detector.  Note that we do not compare our results with the unfolded neutrino spectrum shown in Ref.~\cite{Aartsen:2014gkd}, since it is derived assuming the flavor ratio $1:1:1$.  

Following Refs.~\cite{Gandhi:1995tf,Gandhi:1998ri} (see also, e.g., Ref.~\cite{Connolly:2011vc}), we use the mean inelasticity $\langle y \rangle$ (with modest energy dependence), which is averaged over neutrinos and antineutrinos.  For neutral-current interactions, $E_{\rm dep}\approx \langle y \rangle \epsilon_\nu$ can be used for all the three flavors, and the ratio of the neutral-current cross section to charged-current cross section is approximated to be $\sigma_{\rm NC}/(\sigma_{\rm CC}+\sigma_{\rm NC})\approx0.28$ in this energy range~\cite{Gandhi:1998ri}.  For charged-current interactions of $\nu_e$ and $\bar{\nu}_e$, since both electromagnetic and hadronic cascades contribute to the shower, $E_{\rm dep}\approx \epsilon_\nu$ is expected.  For contained-vertex events caused by $\nu_\mu$ and $\bar{\nu}_\mu$ via charged-current interactions, the energy deposited in the detector mainly comes from hadronic cascades, and we assume $E_{\rm dep}\approx \langle y \rangle \epsilon_\nu$.  Tau leptons are produced by $\nu_\tau$ and $\bar{\nu}_\tau$ via charged-current interactions.  About 64.8\% of tau leptons decay hadronically, so we use $E_{\rm dep}\approx [\langle y \rangle+(2/3)(1-\langle y\rangle)]\epsilon_\nu$ since $2/3$ of the tau lepton energy is used for the second shower.  About 17.8\% of tau leptons decay via $\tau \rightarrow \nu_\tau \bar{\nu}_e e$, for which we use $E_{\rm dep}\approx [\langle y \rangle +(1/3)(1-\langle y \rangle)]\epsilon_\nu$.  About $17.4$\% of tau leptons contribute to track-like events via $\tau \rightarrow \nu_\tau \bar{\nu}_\mu \mu$, for which we use $E_{\rm dep}\approx \langle y \rangle\epsilon_\nu$.  

\begin{figure}[!t]\begin{center}
\includegraphics[width=0.475\textwidth]{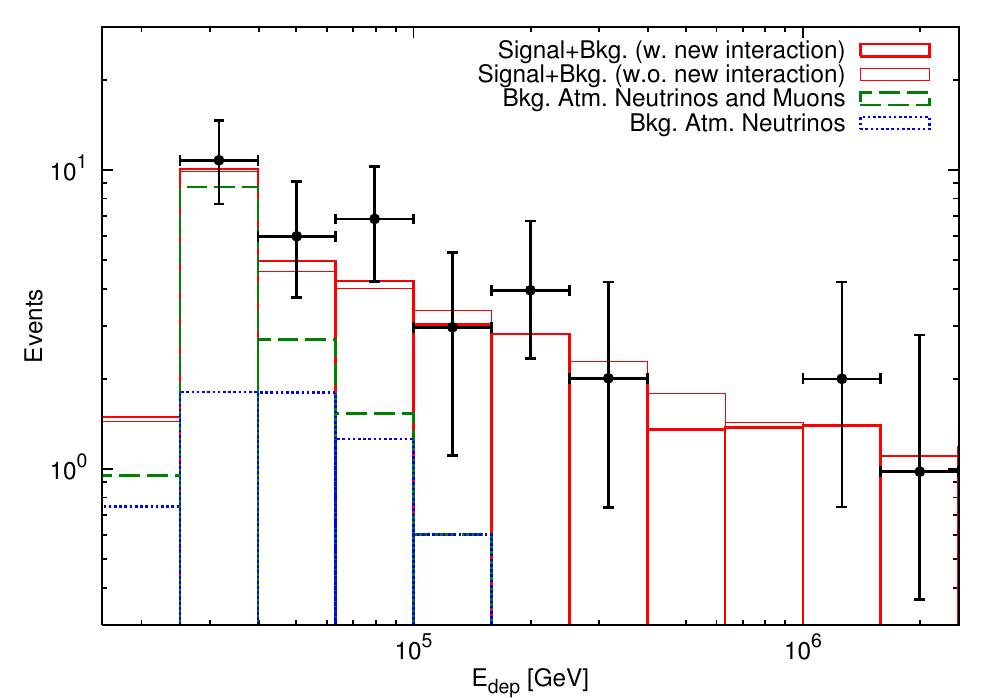}\quad
\includegraphics[width=0.475\textwidth]{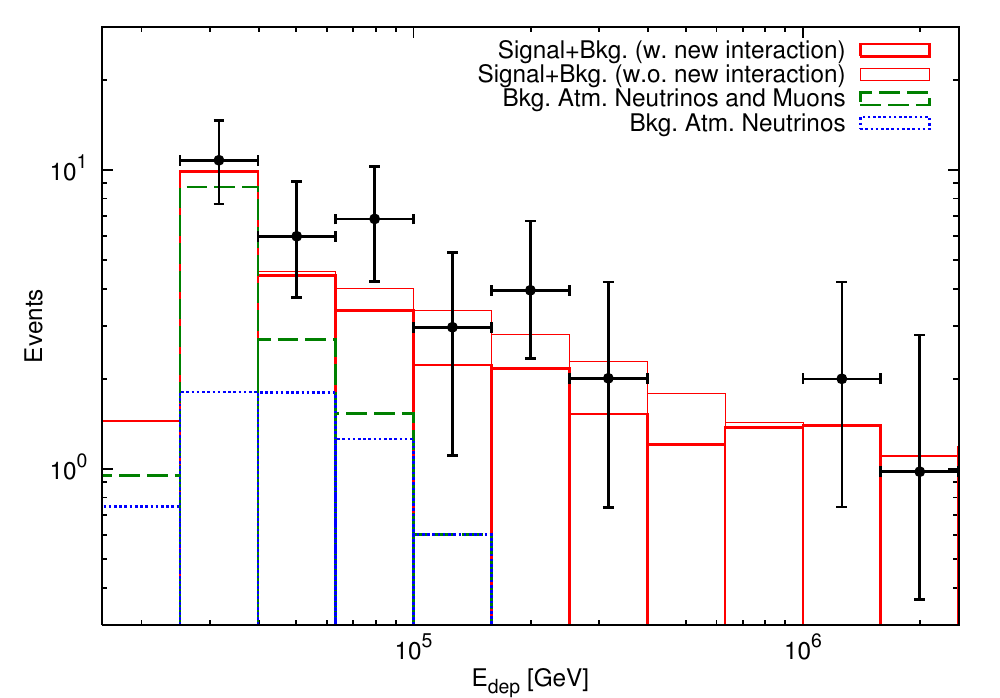}
 \end{center}
\caption{Deposited energy distributions of signals and backgrounds, expected in 988~day observations.  The neutrino events reported by IceCube~\cite{Aartsen:2013bka,Aartsen:2013jdh,Aartsen:2014gkd} are also shown. The atmospheric muon and neutrino backgrounds are taken from Ref.~\cite{Aartsen:2014gkd}.  
We use the same set-up as in Fig.~\ref{fig:Dflav} with $m_\phi=9$~MeV, $\sum_im_{\nu_i}=0.2$~eV and $\G=1.3\times10^{-3}$, assuming normal hierarchy. In the left panel we include regeneration, and in the right panel we do not include regeneration. 
}
\label{fig:NH3yr}
\end{figure}

In Fig.~\ref{fig:NH3yr} we show the results for total (shower-like and track-like) event rates. Our results without the new interactions agree reasonably well with those obtained by the IceCube collaboration~\cite{Aartsen:2014gkd}, for both shower-like and track-like events.   
For the self-interactions, we repeat here the set-up of Fig.~\ref{fig:Dflav} (normal hierarchy), producing a deficit around $E_{\rm dep}\lesssim\epsilon_{\rm res}\sim600$~TeV. 
In the left panel, regeneration is included. Pile-up effects partially cancel the absorption dip, and the resulting observable absorption feature is modest, of order 30\% and covering a narrow range in deposited energy. The pile-up contribution at lower energy is diluted by atmospheric background events. In the right panel regeneration is turned off, reflecting additional decay modes of the exchanged scalar. This leads to an absorption dip that is more pronounced and extending over a wide energy range of order a decade. The  first indication of a signal would most likely be in terms of a deviation from pure power-law spectrum, 
with the spectrum becoming harder above the energy $E_{\rm dep,\,break}\sim\epsilon_{\rm res}$, rather than from the identification of sharp spectral features.  Over all it appears that the no-regeneration case is favorable in terms of detectability.

%%%%%%%%%%%%%%%%%%%%%%%%%%%%
%%%%%%%%%%%%%%%%%%%%%%%%%%%%

\begin{figure}[!t]\begin{center}
\includegraphics[width=0.475\textwidth]{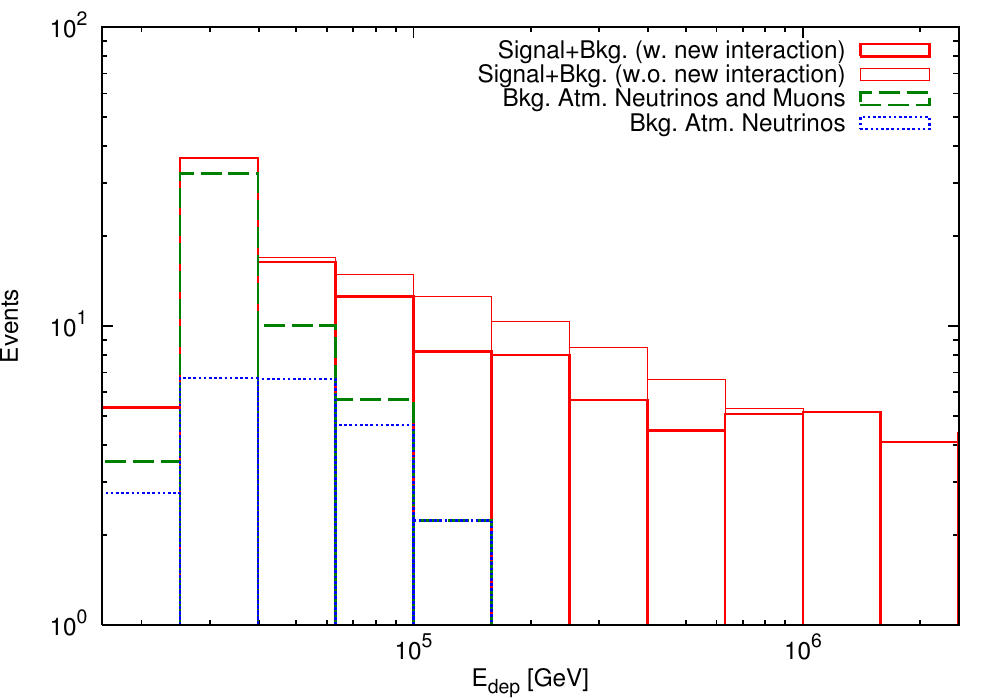}\quad
\includegraphics[width=0.475\textwidth]{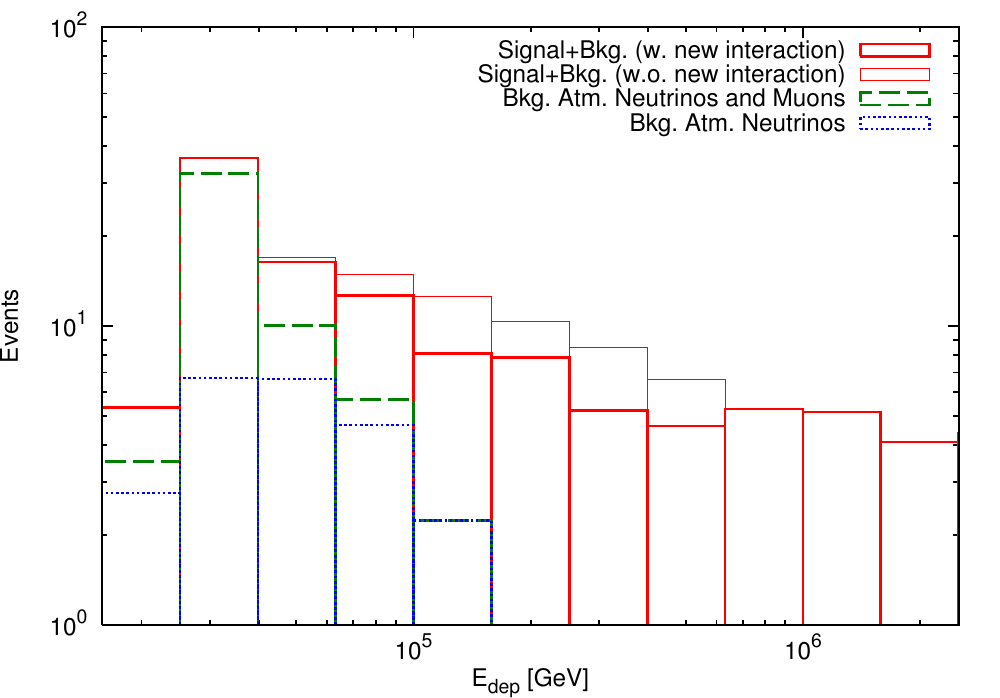}
 \end{center}
\caption{Deposited energy distributions for $m_\phi=5$~MeV, $\sum m_\nu=0.1$~eV, and $\G=10^{-3}$. Ten-year observation is assumed. Normal (inverted) hierarchy is assumed in the left (right) panel.}
\label{fig:NHIH10yr}
\end{figure}

In Fig.~\ref{fig:NHIH10yr} we take $m_\phi=5$~MeV, $\sum_i m_{\nu_i}=0.1$~eV, and $\G=10^{-3}$. Normal (inverted) hierarchy is shown in the left panel (right panel). Here, regeneration is turned off -- as seen above, this is an optimistic scenario that maximizes the visible effect. We illustrate the effect of more statistics, considering ten years of data taking at IceCube. The effect of the different hierarchies, that is very clear in the incoming flux for the value we chose here for $\Sigma_i m_{\nu_i}$ (see Fig.~\ref{fig:Dflav2}), is mostly washed out with the detector response. 
%%%%%%%%%%%%%%%%%%%%%%%%%%%%
%%%%%%%%%%%%%%%%%%%%%%%%%%%%

\begin{figure}[!h]\begin{center}
\includegraphics[width=0.45\textwidth]{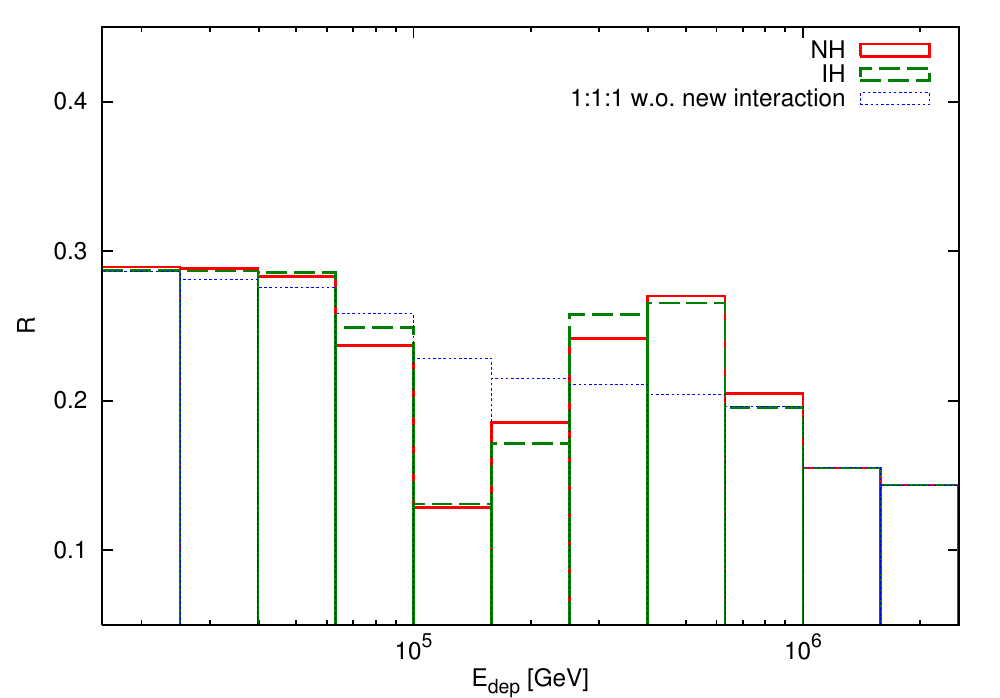}\quad
\includegraphics[width=0.45\textwidth]{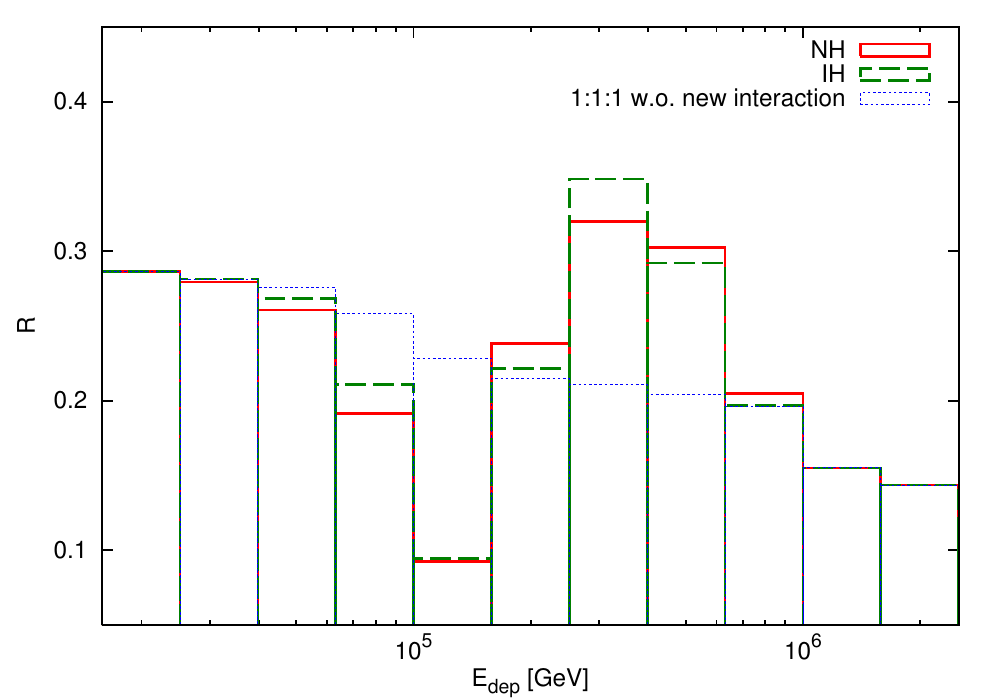}\quad
\end{center}
\caption{The ratio of track-like events to the sum of track-like and shower-like events as a function of $E_{\rm dep}$.  The model parameters are the same as in Fig.~\ref{fig:Dflav}. On the left (right) panel we show the case with (without) regeneration. Only contained events (both cascade and track) are included.
}
\label{fig:Ratio}
\end{figure}

To study the flavor information we introduce the ratio of track-like events to all events, 
\begin{equation}\label{eq:R}
{\mathcal R} \equiv \frac{{\mathcal N}_{\rm track}}{{\mathcal N}_{\rm track}+{\mathcal N}_{\rm shower}}.  
\end{equation}
Note that we still consider only contained events, and $\mathcal R$ in Eq.~(\ref{eq:R}) is defined as a function of $E_{\rm dep}$ rather than $\epsilon_\nu$.  
In Fig.~\ref{fig:Ratio} we show $\mathcal{R}$ vs. $E_{\rm dep}$, using the parameters of Fig.~\ref{fig:Dflav}. In the left (right) panel we consider the case with (without) regeneration.  
We see that $\mathcal R$ is enhanced around $\epsilon_{\rm res}$ compared to the case without self-interactions, since shower-like events are suppressed.  At lower energies, $\mathcal R$ is reduced since track-events with $E_{\rm dep}\approx\langle y\rangle\epsilon_{\rm res}$ are suppressed.   
%%%%%%%%%
%%%%%%%%

\begin{figure}[!t]\begin{center}
\includegraphics[width=0.475\textwidth]{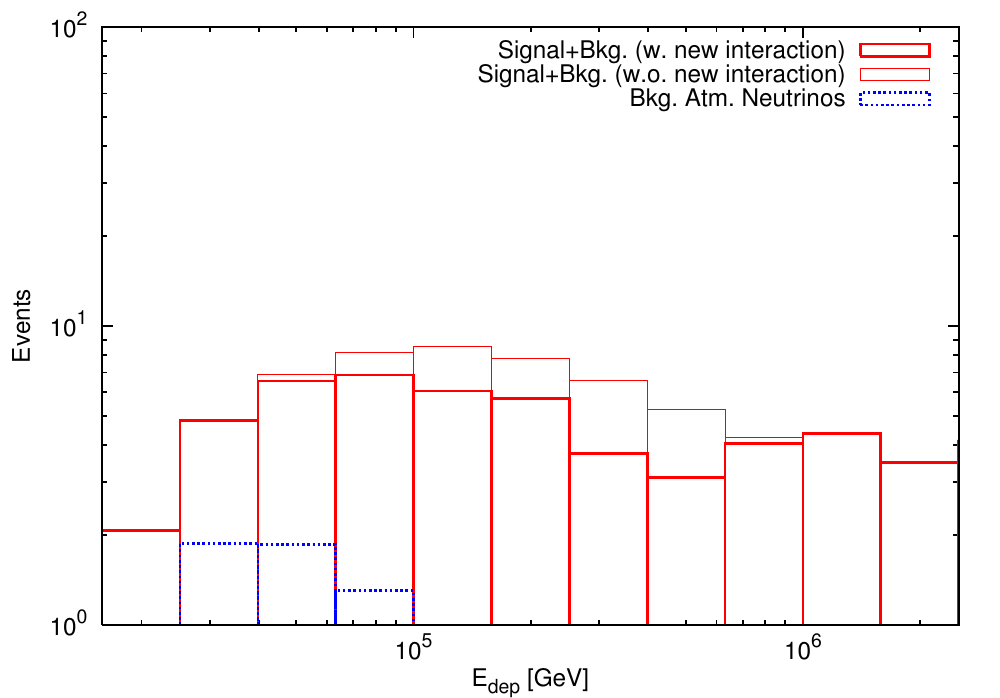}\quad
\includegraphics[width=0.475\textwidth]{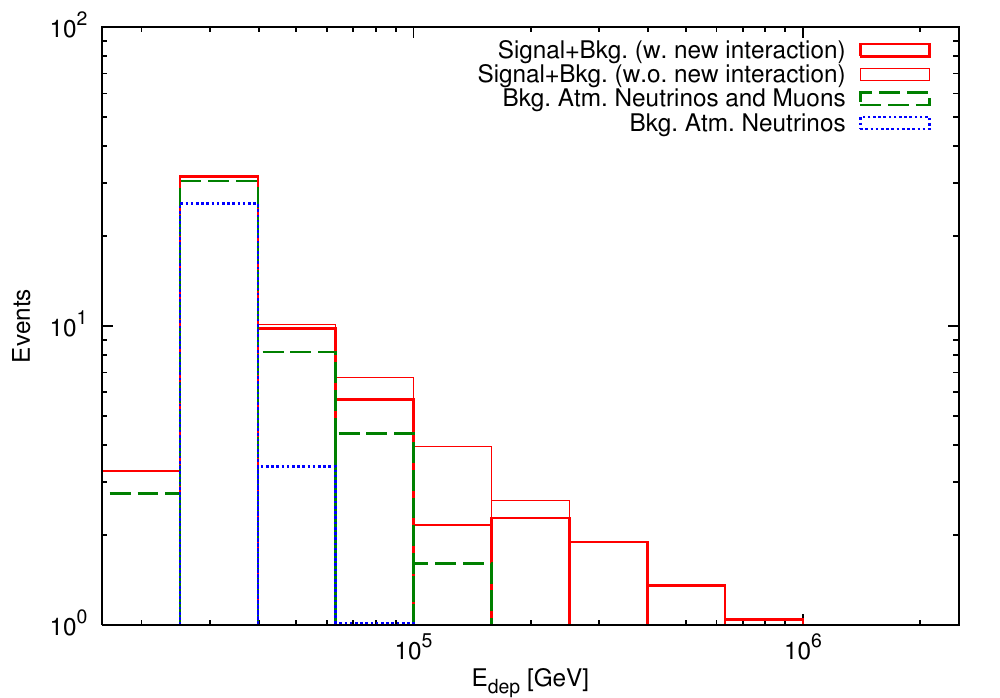}
 \end{center}
\caption{Deposited energy distributions of signals and backgrounds, assuming ten-year exposure.  Shower-like (left) and track-like (right) events are shown separately.  Model parameters are the same as in the right panel of Fig.~\ref{fig:Ratio}.
}
\label{fig:TrackShower10yr}
\end{figure}

While the distribution of $\mathcal{R}$ in Fig.~\ref{fig:Ratio} may look promising, note that it was defined for contained-vertex events only. The main experimental setback here is the low statistics in contained track events, that decrease rapidly at high energy due to the increasing muon penetration length. In Fig.~\ref{fig:TrackShower10yr}, assuming ten-year observations by IceCube, we show deposited energy distributions for shower-like and track-like events in the left and right panels, respectively. The model parameters are the same as in Fig.~\ref{fig:Ratio}, including regeneration. Unfortunately, even in ten years of IceCube exposure, we do not expect to find more than 2-3 contained muon tracks at $E_{\rm dep}\gtrsim200$~TeV. This means that exploiting the variable $\mathcal{R}$ with contained events would require neutrino detectors larger than IceCube.
%%%%%%%%%%%%%%%%%%%%%%%%%%%%
%%%%%%%%%%%%%%%%%%%%%%%%%%%%

\begin{figure}[!t]\begin{center}
\includegraphics[width=0.6\textwidth]{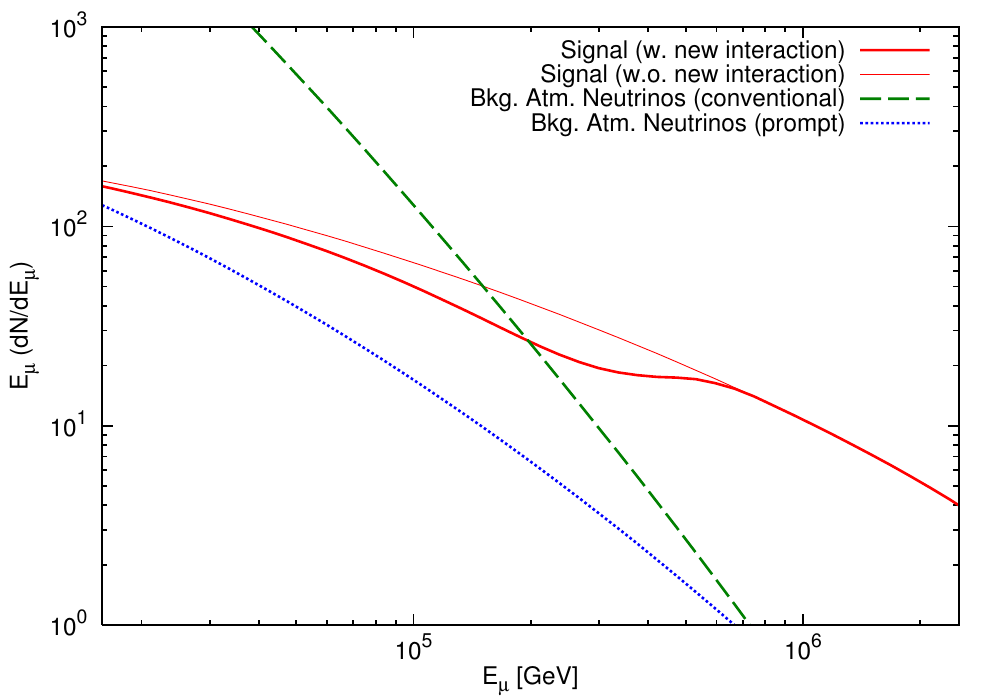}%\quad
 \end{center}
\caption{Through-going muon track energy distribution in IceCube with ten years of exposure. Model parameters are the same as in the  right panel of Fig.~\ref{fig:NHIH10yr}.  The prompt atmospheric neutrino spectrum is taken from Ref.~\cite{Enberg:2008te}.
}
\label{fig:Muon10yr}
\end{figure}

In view of the promising information apparent in Fig.~\ref{fig:Ratio}, the lack of statistics for high energy contained track events strongly motivates the use of through-going muons in the analysis. For through-going muons, we expect comparable statistics at high energy to that of shower events. The neutrino energy resolution and background rejection efficiency, however, are inferior to the contained event case. To demonstrate the potential in a through-going track analysis, in Fig.~\ref{fig:Muon10yr} we plot the distribution of through-going muon track events as function of the muon energy in IceCube after a ten-year exposure, using the same model parameters as in the right panel of Fig.~\ref{fig:NHIH10yr}.  We consider up-going neutrinos since detection of down-going neutrinos suffers from a large atmospheric muon background.  For the calculation we use the method of Ref.~\cite{Laha:2013xua}.  The average muon energy loss is given by $-dE_\mu/dx=\alpha+\beta E_\mu$,  where $\alpha=2\times{10}^{-3}~{\rm GeV}~{\rm cm}^2~{\rm g}^{-1}$ and $\beta=4\times{10}^{-6}~{\rm cm}^2~{\rm g}^{-1}$, and the muon effective area is taken from Ref.~\cite{GonzalezGarcia:2009jc}.  We take into account attenuation of neutrinos during their propagation in the Earth.  %We see that the spectral deficit due to neutrino self-interactions could be observed in a ten-year exposure in IceCube.  Combing analyses for through-going muon and contained-vertex events will be crucial for revealing the flavor ratio as well.  
We leave further analysis details to a dedicated experimental work, but comment that even an energy resolution at the level of a factor of two or so for through-going track events, could add significant information to the interpretation of a signal in IceCube.
%%%%%%%%%%%%%%%%%%%%%%%%%%%%
%%%%%%%%%%%%%%%%%%%%%%%%%%%%

\section{Discussion}\label{sec:disc}

Solar and atmospheric neutrinos are sourced by ``standard astrophysical processes". Nevertheless, modest variations in the spectrum and flavor composition of these neutrinos have taught us a fundamental lesson, that the Standard Model (SM) with massless neutrinos is wrong. Similarly, there is growing, if not yet conclusive, evidence that the high energy extraterrestrial neutrinos detected in IceCube~\cite{Aartsen:2013bka,Aartsen:2013jdh,Aartsen:2014gkd} are coming from astrophysical processes related to the origin of high energy cosmic rays. Still, a fundamental particle physics lesson may also be there to be found.

In this paper we studied a model for low-scale Majorana neutrino mass generation, in which variations in the spectrum and flavor composition of high energy neutrinos could be detected in IceCube.
Our model includes a light scalar, the VEV of which mediates lepton number violation to the SM. As a result, neutrino-neutrino scattering processes involving resonant exchange of the scalar are diagonal in the neutrino mass basis and proportional to powers of neutrino masses.  

We showed that if exchange of the scalar that is responsible for neutrino mass is to produce observable effects in high energy neutrino telescopes, then lepton number violation must be explicit, rather than spontaneously triggered by the scalar. We argued that this requirement is technically natural, and can be implemented in a number of ways. It leads to new phenomenological implications compared to earlier analyses that focused on neutrino mass generation through spontaneous breaking of lepton number~\cite{Chikashige:1980ui,Gelmini:1980re,Riazuddin:1981hz,GonzalezGarcia:1988rw,Chacko:2003dt,Hall:2004yg,Davoudiasl:2005ks,Friedland:2007vv}.

We evaluated the relevant laboratory, astrophysical and cosmological constraints on the model. Significant constraints are found from precision measurements of lepton mixing non-unitarity, leptonic decays of $K$ and $\pi$ mesons and from neutrinoless double-beta decay. These constraints imply that if scalar exchange is relevant at high energy neutrino telescopes, then it must proceed resonantly. A lower bound on the resonance energy, $\epsilon_{\rm res}\gtrsim20$~TeV, is found from the combination of cosmological and laboratory data.

We paid special attention to detector effects, including event topologies and energy resolution at IceCube. Our analysis shows that a significant parameter space exists where neutrino self-interactions could be measured in IceCube in the course of ten years or so.  
%Of course, to make a detection, our understanding of the nature of the astrophysical neutrino source will need to improve. We see no reason to be pessimistic in this respect. 
If indeed neutrino self-interactions arise in relation to a low-scale neutrino mass mechanism, then our work demonstrates that precision laboratory experiments of leptonic flavor physics could provide crucial verification of the model. 
%
%Some of our results are limited to the specific scenario we considered, where neutrino-scalar interactions are diagonal in the mass basis. An important example is the statement that lepton number violation should be explicit in the scalar sector. Other results are more generally applicable, including for instance several of the laboratory bounds we derive on neutrino-scalar interactions. 
It would be exciting to devise a dedicated search strategy in IceCube for anomalous neutrino interactions. The model we propose here provides a well-defined, consistent framework on which to base such a search.

Finally, Refs.~\cite{Ng:2014pca,Ioka:2014kca} used the recent IceCube detection to derive constraints on neutrino self-interactions based on a phenomenological Lagrangian ${\mathcal L}=G\phi\nu\nu$, without specifying a connection to the neutrino mass mechanism. 
Neither of these works considered neutrino flavor effects that, as we showed in this work, are significant in a realistic model. Importantly, neutrino mixing affects the signal observability as spectral features can be washed out directly through the mixing and also by the different detector response to different neutrino flavors.  For the $s$-channel resonance case, this makes the constraints derived in~\cite{Ng:2014pca,Ioka:2014kca} somewhat optimistic.

%%%%%%%%%%%%%%%%%%%%%%%%%
{\it Note added.---}While this work was being completed, we became aware of a related, independent work along similar lines~\cite{Ibe:2014pja}. Our work considers flavor effects in detail and analyzes experimental constraints on the model comprehensively. 

\acknowledgments
We thank Nima Arkani-Hamed, Roni Harnik, Graham Kribs, Paul Langacker, Ranjan Laha, Mehrdad Mirbabayi, Yossi Nir, Shmuel Nussinov, Takemichi Okui, and Kathryn Zurek for useful discussions, and John Beacom, Francis Halzen, Kenny Ng, and Maxim Pospelov for comments on the manuscript. KB and AH were supported by the DOE grant de-sc000998. KM was  supported by NASA through Hubble Fellowship, Grant No. 51310.01 awarded by the STScI, which is operated by the Association of Universities for Research in Astronomy, Inc., for NASA, under Contract No. NAS 5- 26555. KB and KM thank Sheldon Campbell and Carsten Rott for organizing the CCAPP workshop on ''Cosmic Messages in Ghostly Bottles: Astrophysical Neutrino Sources and Identication," where this work was initiated. The work of KB and AH was supported in part by the NSF under Grant No. PHYS-1066293 and the hospitality of the Aspen Center for Physics.

%%%%%%%%%%%%%%%%%%%%%%%%%%%%%%%%%%% 
\begin{appendix}
%

%%%%%%%%%%%%%%%%%%%%%%%%%%%%%%%%
\section{Neutrino scattering cross sections}\label{app:xs}

Consider the cross section for $\nu_i(\epsilon)\nu_i(m_{\nu_i})\to \nu_i\nu_i$, representing the scattering of a high energy Majorana neutrino of energy $\epsilon$ off of a background neutrino at rest. Denote the energy of one of the final state neutrinos by $x\epsilon$ with $0<x<1$.\footnote{The energy of the other neutrino is $(1-x)\epsilon$. We note that the expression for the cross section used in Ref.~\cite{Ioka:2014kca} is reproduced if we consider $\phi$ as a real scalar field and use $n(z=0)\simeq56$~cm$^3$ as the number density of target C$\nu$B, restricting to the negative helicity states that participate in neutrino-neutrino scattering.} Exchange of $\phi$ yields
\be\label{eq:sig1}\frac{d\sigma_{ii\to ii}[\hat s,x]}{dx}&=&\left|\G_{i}\right|^4\frac{\overline{\left|\mathcal{M}_{ii\to ii}\right|^2}}{16\pi \hat s},\ee
with the spin-averaged matrix element
\be \overline{\left|\mathcal{M}_{ii\to ii}\right|^2}&=&\frac{1}{2}\left(\frac{\hat s^2}{(\hat s-m_\phi^2)^2+m_\phi^2\Gamma_\phi^2}+\frac{\hat t^2}{(\hat t-m_\phi^2)^2+m_\phi^2\Gamma_\phi^2}+\frac{\hat u^2}{(\hat u-m_\phi^2)^2+m_\phi^2\Gamma_\phi^2}\right).\ee
Here, the Mandelstam variables are $\hat s=2m_{\nu_i}\epsilon$, $\hat t=-(1-x)\hat s$, and $\hat u=-x\hat s$. 
We assume that the scalar sector is perturbative with $\Gamma_\phi\ll m_\phi$, and only keep insertions of $\Gamma_\phi$ in the denominator. 

Another process of interest is the scattering $\nu_i(\epsilon)\nu_i(m_{\nu_i})\to \nu_j\nu_j$ with $i\neq j$. The cross section for this process is given by
\be\label{eq:sig2}\frac{d\sigma_{ii\to jj}[\hat s,x]}{dx}&=&\left|\mathcal{G}_{i}\right|^2\left|\mathcal{G}_{j}\right|^2\frac{\overline{\left|\mathcal{M}_{ii\to jj}\right|^2}}{16\pi \hat s},\;\;\;\;\;{\rm with}\;\;\;\;\; \overline{\left|\mathcal{M}_{ii\to jj}\right|^2}=\frac{1}{2}\frac{\hat s^2}{(\hat s-m_\phi^2)^2+m_\phi^2\Gamma_\phi^2},\ee
and the Mandelstam variables are as above. 

Lastly we have $\nu_i(\epsilon)\nu_j(m_{\nu_i})\to \nu_i\nu_j$. Choosing $x\epsilon$ to be the energy of the final state $\nu_i$, we find
\be\label{eq:sig3}\frac{d\sigma_{ij\to ij}[\hat s,x]}{dx}&=&\left|\mathcal{G}_{i}\right|^2\left|\mathcal{G}_{j}\right|^2\frac{\overline{\left|\mathcal{M}_{ij\to ij}\right|^2}}{16\pi \hat s},\;\;\;\;\;{\rm with}\;\;\;\;\; \overline{\left|\mathcal{M}_{ij\to ij}\right|^2}=\frac{1}{2}\frac{\hat t^2}{(\hat t-m_\phi^2)^2+m_\phi^2\Gamma_\phi^2},\ee
and the Mandelstam variables are $\hat s=2m_{\nu_j}\epsilon$, $\hat t=-(1-x)\hat s$, and $\hat u=-x\hat s$.

\begin{figure}[!h]\begin{center}
\includegraphics[width=0.6\textwidth]{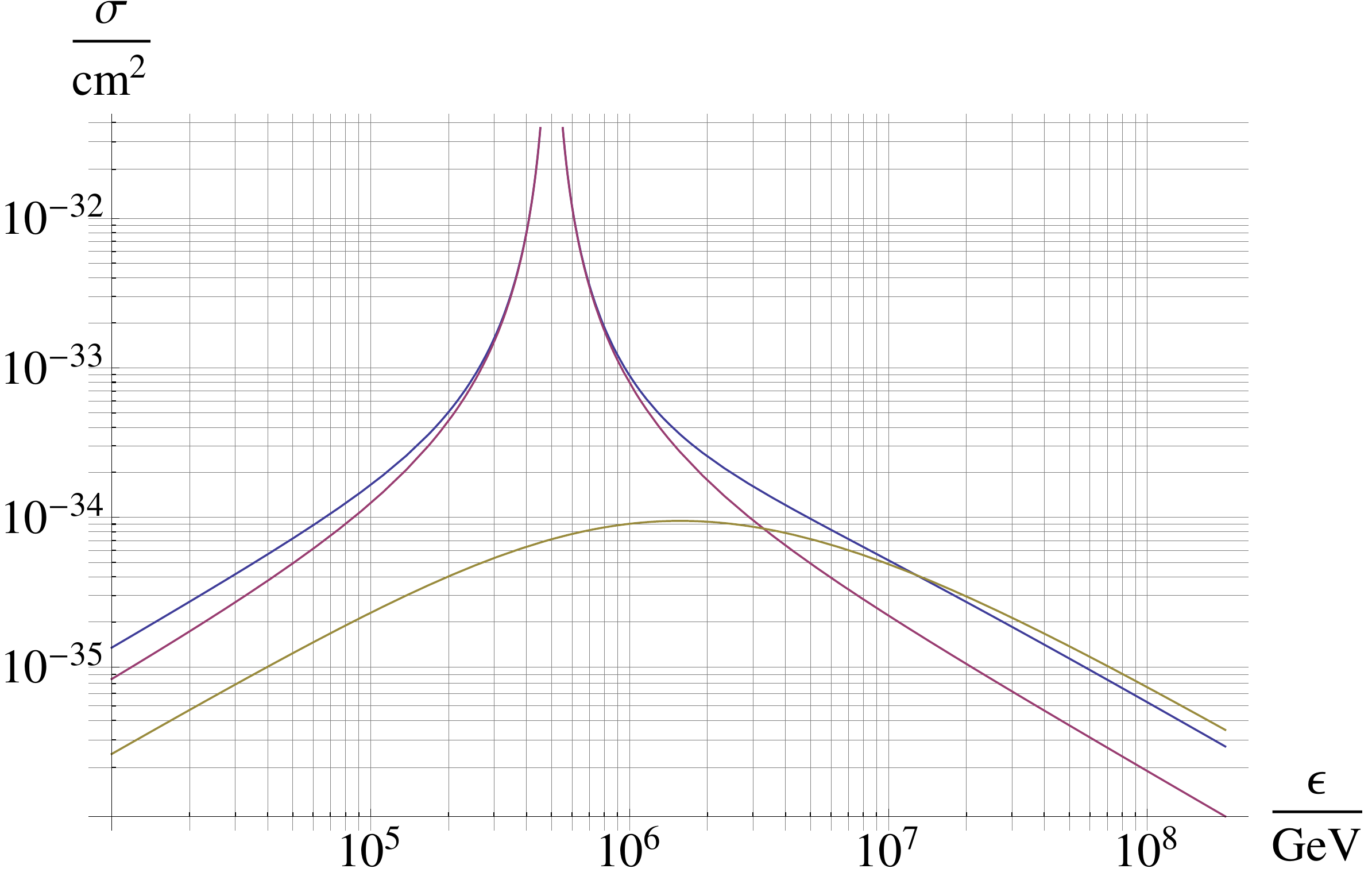}\end{center}
\caption{Total cross sections for $ii\to ii$ (blue), $ii\to jj$ (purple), and $ij\to ij$ (brown), with parameters $m_\phi=10$~MeV, $\Gamma_\phi=10^{-4}m_\phi/(4\pi)$, $m_{\nu_i}=2m_{\nu_j}=0.1$~eV, and $\G_{i}=\G_{j}=10^{-2}$.
}
\label{fig:xs}
\end{figure}

We denote the total cross sections by
\be\sigma_{X}[\hat s]=c_X\int_0^1dx\frac{d\sigma_X[\hat s,x]}{dx},\ee
where $c_X=1/2$ in the case of two identical particles in the final state ($ii\to ii,\;ii\to jj$) and $c_X=1$ for $ij\to ij$ with $i\neq j$. 
In Fig.~\ref{fig:xs} we plot the total cross sections for the reactions above, using $m_\phi=10$~MeV, $\Gamma_\phi=10^{-4}m_\phi/(4\pi)$, $m_{\nu_i}=2m_{\nu_j}=0.1$~eV, and $\G_{i}=\G_{j}=10^{-2}$.
%%%%%%%%%
%%%%%%%%

The contribution of the s-channel diagrams above depends crucially on the decay width of the exchanged scalar. This can be computed if no other decay paths except for the two-neutrino state exist,
\be\label{eq:Gs}\Gamma_\phi&=&\frac{m_\phi}{32\pi}\sum_i\left|\G_{i}\right|^2.\ee

In the scattering calculations above, we summed scalar and pseudo-scalar exchange diagrams, ignoring the small mass splitting between these states. We now comment on the breaking of scalar--pseudo-scalar mass degeneracy due to the explicit breaking of lepton number in the model. Corrections to the near-degeneracy of the scalar ($s$) and pseudo-scalar ($a$) components of $\phi=(s+ia)/\sqrt{2}$ arise as $\Delta m_\phi^2=m_s^2-m_a^2=2\lambda_\phi\,\mu^2=\frac{2\lambda_\phi}{\G^2}m_{\nu}^2$. This splitting means that scalar and pseudo-scalar s-channel diagrams go resonant at slightly different neutrino energy, $(\epsilon_{{\rm res},s}-\epsilon_{{\rm res},a})/\epsilon_{\rm res}=\Delta m_\phi^2/m_\phi^2$, where $\epsilon_{\rm res}$ denotes the mean resonance energy. This should be compared to the width of each resonance, caused by the decay width of the states, $\Delta\epsilon_{\rm res}/\epsilon_{\rm res}=\Gamma_\phi/m_\phi$. In the parameter space of interest to us ($m_\phi\gtrsim$~MeV, $\G\gtrsim10^{-3}$) and for reasonable values of $\lambda_\phi\lesssim0.1$, we see that the mass splitting is smaller than the width of the states, and can be ignored: $(\epsilon_{{\rm res},s}-\epsilon_{{\rm res},a})/\epsilon_{\rm res}=\frac{2\lambda_\phi}{\G^2}\frac{m_{\nu}^2}{m_\phi^2}\ll\Delta\epsilon_{\rm res}/\epsilon_{\rm res}\sim\frac{\G^2}{32\pi}$.

%%%%%%%%%%%%%%%%%%%%%%%%%%%%%%%
\section{Experimental constraints}\label{app:expcst}

Experimental constraints on $\nu\nu$ interactions were considered in, e.g.,~\cite{Bardin:1970wq,Bilenky:1999dn,Bilenky:1992xn,Gavela:2008ra,Gelmini:1982rr,Lessa:2007up}, some of which allowed for a light mediator and some took an effective theory approach. Below we recalculate the most relevant constraints, finding that the strongest generic bounds on $\G$ come from kaon decays, independent of the scalar mass for $m_\phi\ll m_K$ as is relevant for this work. Stronger bounds exist from neutrinoless double-beta decay, but apply only for a light scalar $m_\phi<2$~MeV. 
Strong constraints, though specific to our model with heavy sterile neutrinos, are found  from PMNS matrix non-unitarity, and apply regardless of the interactions of $\phi$.

\paragraph{Light meson decays.} The decay mode $\pi^+\to e^+\nu\phi$ opens the possibility for pion decay into an electron with no helicity suppression~\cite{Gelmini:1982rr,Lessa:2007up}. In the limit $m_\phi\ll m_\pi$ we find, in agreement with~\cite{Lessa:2007up} 
\be\label{eq:Pidec} \frac{BR(\pi^+\to e^+\nu\phi)}{BR(\pi^+\to e^+\nu_e)}&=&\frac{\sum_i\G_i^2|U_{ei}|^2}{12(4\pi)^2}\frac{m_\pi^2}{m_e^2}.
\ee
Here $U_{\alpha i}$ is the PMNS matrix.  
Ref.~\cite{Beringer:1900zz} gives $BR(\pi^+\to e^+\nu_e)=(1.230\pm0.004)\times10^{-4}$ and $BR(\pi^+\to e^+\nu_e\nu\bar\nu)<5\times10^{-6}$ @ 90\%CL. Imposing that our three-body mode proceeds at a rate smaller than the quoted error on the two-body $\pi^+\to e^+\nu$, leads to the conservative bound $\sum_i\G_i^2|U_{ei}|^2\lesssim10^{-4}$. This bound is conservative, as it does not take into account experimental analysis cuts on the invariant mass distribution of the charged lepton decay product and the initial meson state. It is plausible that the constraint on $\pi^+\to e^+\nu\nu\bar\nu$ is more directly applicable to our model, in which case we have $\sum_i\G_i^2|U_{ei}|^2\lesssim10^{-3}$. 

A stronger bound is found from $K$ decays. Ref.~\cite{Lazzeroni:2012cx} quotes the experimental result $R_K^{({\rm exp})}\equiv\Gamma(K^\pm\to e^\pm\nu)/\Gamma(K^\pm\to \mu^\pm\nu)=(2.488\pm0.010)\times10^{-5}$. The SM prediction~\cite{Cirigliano:2007xi} is $R_K^{({\rm SM})}=(2.477\pm0.001)\times10^{-5}$. Ignoring  again the analysis cuts and assuming $m_\phi\ll m_K$, a conservative bound can be put using
\be\label{eq:Kdec}\frac{\delta R_K}{R_K}\approx\frac{\sum_i\G_i^2|U_{ei}|^2}{12(4\pi)^2}\frac{m_K^2}{m_e^2}
\ee
and imposing $\delta R_K/R_K\lesssim10^{-2}$.

Considering four-lepton decays, the PDG~\cite{Beringer:1900zz} quotes $BR(K^+\to \mu^+\nu_\mu\nu\bar\nu)<6\times10^{-6}$ @ 90\%CL. We impose the same upper limit on $BR(K^+\to \mu^+\nu\phi)$ and write 
\be\label{eq:Kdec2} \frac{BR(K^+\to \mu^+\nu\phi)}{BR(K^+\to \mu^+\nu)}=\frac{\sum_i\G_i^2|U_{\mu i}|^2}{12(4\pi)^2}\frac{m_K^2}{m_\mu^2}
\ee
with $BR(K^+\to \mu^+\nu)\approx0.64$. Here, using the prescription given in Ref.~\cite{Pang:1989ut}, we also verify the effect of analysis cuts. Folding the muon spectrum in the decay $K\to\mu\nu\phi$ with the efficiency function of~\cite{Pang:1989ut} we obtain $BR(K^+\to \mu^+\nu\phi)<7.6\times10^{-6}$ @ 90\%CL, a slightly weaker bound than what would naively be deduced from~\cite{Beringer:1900zz}. We comment that Ref.~\cite{Laha:2013xua} derived much stronger limits for the case of neutrino couplings to a light vector boson.  

\begin{figure}[!h]\begin{center}
\includegraphics[width=0.75\textwidth]{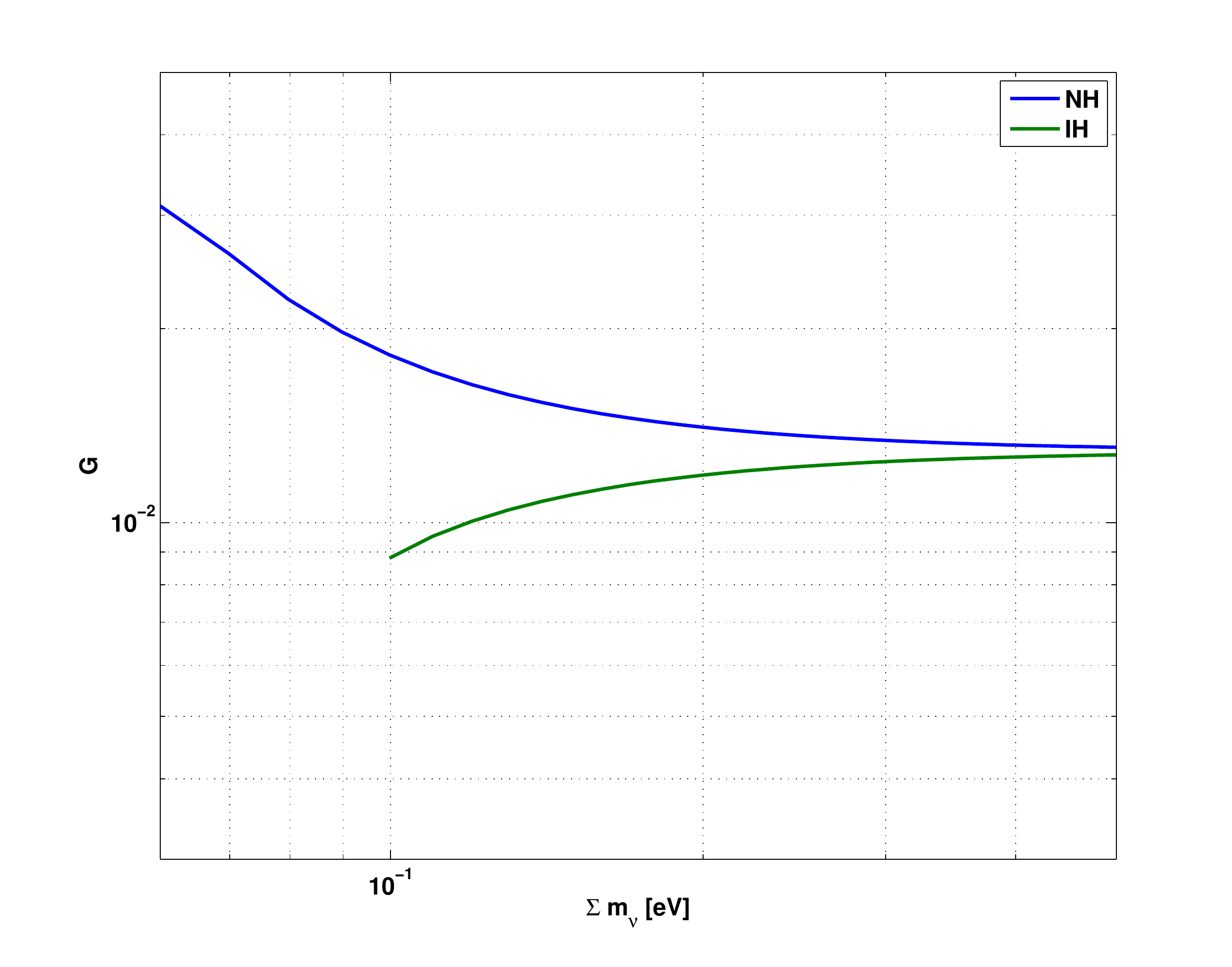}\end{center}
\caption{Upper bound on $\G=\sum_i\G_i$ from kaon decay, using numerical values for the neutrino oscillation parameters from~\cite{Capozzi:2013csa}. The x-axis specifies the sum of neutrino masses. The blue (green) curve denotes the normal (inverted) mass hierarchy. Comparable but somewhat weaker constraints are found from pion decay.}
\label{fig:Kbound}
\end{figure}

Using the measured PMNS matrix elements, we have from Eqs.~(\ref{eq:Kdec}-\ref{eq:Kdec2}),
\be\label{eq:Kdecbounde}\sum_i\G_i^2|U_{e i}|^2\approx0.77\,\G_1^2+0.29\,\G_2^2+0.02\,\G_3^2&<&2\times10^{-5},\\
\label{eq:Kdecboundmu}\sum_i\G_i^2|U_{\mu i}|^2\approx0.25\,\G_1^2+0.28\,\G_2^2+0.48\,\G_3^2&<&8\times10^{-4}.
\ee
Given a value for the sum of neutrino masses, our definition of $\G=\sum_i\G_i$ (see Eq.~(\ref{eq:Gdef})) can be used together with the known neutrino mass differences to translate Eqs.~(\ref{eq:Kdecbounde}-\ref{eq:Kdecboundmu}) into a constraint on $\G$. This constraint is depicted in Fig.~\ref{fig:Kbound}.
%%%%%%%%%
%%%%%%%%
We comment that comparable but somewhat weaker constraints can be derived from the charged lepton spectrum and total rates of muon and tau decays~\cite{Lessa:2007up} .

\paragraph{Neutrinoless double-beta decay.} 
In our model, neutrinoless double-beta decay ($0\nu\beta\beta$) can occur with additional scalar emission replacing the Majorana mass insertion, $(A,Z)\to(A,Z+2)+2e+\phi$. For $\phi$ much lighter than about 1~MeV the constraint is very strong; for example, Refs.~\cite{Gando:2012pj,Arnold:2013dha} quote bounds equivalent to $\sum_i\G_i^2|U_{ei}|^2\lesssim10^{-10}$ and $1.6\times10^{-9}$, respectively, orders of magnitude stronger than the bound from meson decay. However, the $0\nu\beta\beta$ bound becomes weaker due to kinematic suppression as the mass of $\phi$ approaches the Q-value of the reaction, and for $m_\phi> Q-2m_e$ the decay is kinematically blocked. 

Ref.~\cite{Barabash:2014uqa} surveyed leading $0\nu\beta\beta$ constraints. From their list, Ref.~\cite{Arnold:2013dha} using $^{100}$Mo with $Q\approx3.03$~MeV has the highest Q-value (the Q-value for~\cite{Gando:2012pj}, using $^{136}$Xe, is $Q\approx2.5$~MeV). Using the result of~\cite{Arnold:2013dha} and taking into account neutrino mixing, we obtain the bound
\be\label{eq:0nubb}\G\lesssim10^{-4},\;\;\;\;\;\;{\rm valid\;for\;\;\;\;m_\phi<2~MeV.}\ee
This bound is conservative as it does not take into account the kinematic suppression near threshold.

\paragraph{$Z$ invisible width.} 
Precision $Z$-pole data constrains our light scalar as it corrects the decay $Z\to\nu\nu$ at loop level and adds the channel $Z\to\nu\nu\phi$ at tree level. We calculated both of these effects finding that the constraints are weaker than the pion and kaon constraints by about an order of magnitude.

\paragraph{$\mu\to e\gamma$.}
Our model is similar to the inverse seesaw model in that it lowers the scale of sterile neutrinos down to $\mathcal{O}(10~{\rm TeV})$. In order to avoid excessive lepton flavor violation at low energies, the flavor structure of the various couplings in Eq.~(\ref{eq:Luv0}) cannot be arbitrary. A simple possibility is to assume that lepton flavor violation is only mediated through the couplings of $\Phi$. In the language of Eq.~(\ref{eq:Luv0}), we take $M$ and $y$ to be proportional to the unit matrix in lepton flavor, while the coupling $y'$ possesses the structure in lepton flavor needed to reproduce the observed PMNS matrix and the neutrino mass spectrum. In this set-up, given the smallness of the VEV $\langle\Phi\rangle=\mu$ in our framework, the branching fraction $BR(\mu\to e\gamma)$ is negligible.

\paragraph{Non-unitary leptonic mixing.}
Mixing of the active neutrinos with the heavy sterile states in our model results with an effective active neutrino mixing matrix $U$ that is non-unitary (see, e.g.,~\cite{Weiland:2013wha,Antusch:2008tz}). Very recently, Ref.~\cite{Antusch:2014woa} reported constraints on non-unitary neutrino mixing, derived by combining a long list of precision flavor and electroweak laboratory measurements. 
Parametrizing the non-unitarity by $UU^\dag=1+\epsilon$, where $\epsilon$ is a matrix in lepton flavor, Ref.~\cite{Antusch:2014woa} finds $-2.1\times10^{-3}<\epsilon_{ee}<-2\times10^{-4},\;-4\times10^{-4}<\epsilon_{\mu\mu}<0,\;-5.3\times10^{-3}<\epsilon_{\tau\tau}<0$ at 90\%CL. 

For our model, using Eq.~(\ref{eq:effG}) we find $\epsilon=-\frac{v^2}{2}(M^{-1}y)^\dag M^{-1}y=-\sqrt{\epsilon}^\dag\sqrt{\epsilon}$, defining the matrix $\sqrt{\epsilon}=(v/\sqrt{2})M^{-1}y$. We also have $\G=\sqrt{\epsilon}^T y'\sqrt{\epsilon}=m_\nu/\mu$. Therefore, the constraints on PMNS matrix non-unitarity translate in our model to constraints on the coupling $y'$, where for fixed $\G$ an upper bound on $\epsilon$ leads to a lower bound on $y'$. The detailed constraints depend on how lepton flavor violation occurs in the model. Assuming that $y$ and $M$ are diagonal and universal,  $\epsilon$ and $\sqrt{\epsilon}$ are also diagonal and universal. With some abuse of notation we can write $y'=-\G/\epsilon$, thinking of $\epsilon$ as a number. 
We learn that the analysis of~\cite{Antusch:2014woa} puts strong constraints on the model. In order to have $\G\gtrsim10^{-3}$ we need to require sizable couplings,  $y'=\mathcal{O}(1)$. In the body of the paper, in the examples of Figs.~\ref{fig:Dflav}-\ref{fig:Dflav2} and in Sec.~\ref{ssec:detect} we took care to maintain $\G_i<5\times10^{-4}$, consistent with the constraints on PMNS non-unitarity at the $\sim$90\%CL for ${\rm max}\{y'\}\approx1$ and for the simple flavor structure assumed here.

\paragraph{Astrophysical and cosmological constraints.}
A light $m_\phi$ would add to the number of relativistic degrees of freedom during big-bang nucleosynthesis (BBN). As we saw, however, neutrinoless double-beta decay already implies $m_\phi>2$~MeV, making the BBN constraint irrelevant. Constraints from the observation of a neutrino burst from SN1987A (see, e.g.~\cite{Kolb:1987qy}) are also weaker than the laboratory constraints derived above.  
Refs.~\cite{Cyr-Racine:2013jua,Archidiacono:2013dua} derived constraints on neutrino self interactions due to the effect on the CMB anisotropies (see also~\cite{Friedland:2007vv} for discussion and an earlier forecast). Their analysis implies $\G\lesssim10^{-1}\left(\frac{m_\phi}{10~{\rm MeV}}\right)$, somewhat weaker than laboratory constraints for most values of $m_\phi$ of interest to us.

\end{appendix}
%============================================================================================================================

\bibliography{ref}

\end{document}